\documentclass[alpha-refs]{wiley-article}
\usepackage{lineno}

\usepackage{color}
\usepackage{ulem}

\usepackage{siunitx}
\usepackage{graphicx}
\papertype{Original Article}
\paperfield{Journal Section}

\title{Observations and numerical simulations of a valley-exit wind in the Alpine Bolzano basin}

\author[1,2]{Federica Gucci}
\author[2,3]{Andrea Zonato}
\author[2,4]{Marco Falocchi}
\author[2,5]{Dino Zardi}
\author[2]{Lorenzo Giovannini}

\affil[1]{Institute of Geophysics and Meteorology, University of Cologne, Cologne, Germany}
\affil[2]{Department of Civil, Environmental and Mechanical Engineering, University of Trento, Trento, Italy}
\affil[3]{CIMA Research Foundation, Savona, Italy}
\affil[4]{CISMA S.r.l., Bolzano, Italy}
\affil[5]{Center Agriculture Food Environment (C3A), University of Trento, Trento, Italy}

\corraddress{F. Gucci, Institute of Geophysics and Meteorology, University of Cologne, Cologne, Germany.\\ Email: fgucci@uni-koeln.de\\L. Giovannini, Department of Civil, Environmental and Mechanical Engineering, University of Trento, Trento, Italy.\\ Email: lorenzo.giovannini@unitn.it}
\runningauthor{Gucci et al.}

\begin{document}

\maketitle

\begin{abstract}
The characteristics of the nocturnal drainage wind flowing from the tributary Isarco Valley into the Bolzano basin, in the Italian Alps, during wintertime are investigated. Analyses are performed by combining measurements from an intensive field campaign and the output of four high-resolution numerical simulations, run with the Weather Research and Forecasting (WRF) model using different planetary boundary-layer (PBL) schemes. Two episodes are identified, based on the vertical temperature stratification in the basin and the evolution of the drainage flow at the valley exit. Numerical results show that the drainage flow behaves as a valley-exit wind, whose main structure at the exit of the valley is well captured by the model independently of the PBL scheme. However, the model struggles to correctly reproduce the temperature stratification in the basin, with better results when a PBL scheme including, among others, a prognostic equation for the temperature variance and a counter-gradient term is used. This has an impact on the simulation of the onset and duration of the valley-exit wind, which are sensitive to the temperature contrasts between the valley and the basin. Overall, the model is able to reproduce the different behavior of the drainage wind at the exit of the valley in the two case studies. It is found that the presence of a cold air pool in the basin favors an upward trajectory of the flow at the exit of the valley, resulting in unperturbed calm wind conditions in the lower levels. On the other hand, with weak temperature stratification, the drainage flow closely follows the topography, resulting in strong winds also near the surface.
\end{abstract}
\section{Introduction}

Almost half of the Earth's land surface is covered by hilly or mountainous terrain \citep{Meybeck2001}. Hence, the understanding of atmospheric circulations over these regions is an important issue for many sectors, including weather forecasting, pollutant dispersion and renewable energy source assessment. 
However, atmospheric circulations over mountainous regions are much more complex than over flat terrain, due to the interaction between the flow and the orography at different spatial and temporal scales \citep{Lehner2018, Adler2021, Babic2021, goger2025scale}. 

Under fair weather conditions, different types of thermally-driven circulations typically develop over mountain regions \citep{Zardi2013}, driven by pressure gradients produced by differential heating of the atmosphere in adjacent areas. In particular, mountain valleys are affected by the development of cross- and along-valley circulations, generally called slope and valley wind systems, respectively \citep{Whiteman1990,Rucker2008, Schmidli2009, Giovannini2017}.

When the surface energy budget results in radiative loss, the near-surface atmospheric layer becomes stably stratified, reducing vertical motion and turbulent mixing. 
Thermally-driven down-valley and down-slope flows may contribute, through convergence from tributary valleys and sidewalls, to strengthen the stratification of the valley atmosphere, with the development of strong thermal inversions and cold-air pool (CAP) episodes \citep{DeWekker2006}. CAPs consist of a topographically confined, stagnant layer of air close to the surface colder than the atmosphere above \citep{Whiteman2001}. CAPs typically form in local depressions and valley constrictions \citep{Clements2003, Lundquist2008} during the evening/night and break up after sunrise of the next day, due to the energy input provided by solar radiation and the consequent surface heating. However, especially during wintertime and in mountain areas, the CAP can persist for several days, due to the weaker solar radiation input, also determined by mountain shading \citep{Clements2003, Lundquist2008, Daly2009, Lareau2013, Rauchocker2024}.  

There is a mutual interaction between the local airflow and the CAP. Convergence of down-slope flows contributes to increasing the depth of the CAP until the down-valley flow is fully developed. Once a CAP has formed, it may, in turn, influence the local circulation. A deep stable layer, in fact, may act as a material surface (sometimes referred to as \textit{effective mountain}) to the airstream impinging on it \citep{Neff1987,Neff1989,Lin2005,Reeves2006, Munoz2020}. 
Numerical simulations have shown that down-slope winds may split when they reach the cold air accumulated inside a basin, partly flowing close to the surface and partly above the CAP \citep{Cuxart2007}. On the other hand, the presence of flows on top of the CAP may contribute to its erosion by shear-driven turbulence at the interface \citep{2015_Lareau_erosion}, a process that was shown to be non-negligible especially for CAP erosion by f\"ohn \citep{2005_Jaubert, 2022_Haid_sheardrivenerosion}. 

The present work focuses on a nocturnal drainage current flowing from a tributary valley into the basin of Bolzano (Italian Alps) during wintertime. The formation of a CAP in the basin is expected to alter the spatial characteristics of the current, which are investigated under two distinct temperature stratifications in the basin, combining measurements from an intensive field campaign and results from numerical simulations with the Weather Research and Forecasting (WRF) model. Meteorological data were collected during the Bolzano Tracer EXperiment (BTEX) in January and February 2017 \citep{Tomasi2019,Falocchi2020,Zardi2021}, providing both surface measurements and vertical profiles of temperature and wind speed and direction. These experimental data serve both to characterise local meteorological conditions, in particular the drainage wind, and to evaluate model results. 
The representation of very stable planetary boundary layers (PBL) with numerical models is challenging, due to the reduced and intermittent turbulent mixing \citep{Sandu2013, Boyko2024}. This, combined with the complex topography where CAP generally occurs and the many atmospheric processes that shape its evolution and structure \citep{2011_Lareau_CAPprocesses}, makes it even more difficult to represent such a phenomenon \citep{2021_Arthur_CAPhordiff, 2023_Adler_CAPCloud} and makes the model more sensitive to the representation of the PBL dynamics \citep{2024_Ma_CAPsimPBLs}. For this reason, the performance of four different PBL schemes in reproducing the evolution of the drainage flow once it reaches the basin is assessed. Among the PBL schemes tested, a new scheme including a prognostic equation for turbulent potential energy (TPE) is adopted, since it is known that the conversion between turbulent kinetic energy (TKE) and TPE is a crucial mechanism in stable boundary layer dynamics \citep{Zilitinkevich2007}. While this new scheme has been evaluated in idealised simulations of mountain environments \citep{Zonato2022}, it is tested for the first time here in real-case complex-terrain scenarios.

Based on model verification results, a simulation for each episode is then selected to provide a more complete analysis of the spatial development of the drainage flow and its interaction with the air in the Bolzano basin, complementing the punctual information from the experimental measurements. The first insights from the BTEX experiment had shown that the nocturnal drainage current presents characteristics similar to valley-exit jets \citep{Tomasi2019} and the numerical simulations will be used to confirm this hypothesis. Also, the interaction of CAP events with strong flow is a complex process that cannot be captured fully by point measurements. Therefore, the results of the numerical simulations are used to understand how the formation of a CAP affects the spreading of the valley-exit wind into the basin. While several observational and numerical studies have been conducted to investigate CAP processes \citep{Clements2003, 2011_Price_COLPEX, Arduini2016, Rauchocker2024}, only a few have analysed the interaction with the local circulation, mainly focusing on f\"ohn events \citep{2006Flamant, Haid2020, 2021Umek}, rather than smaller-scale flows, such as valley-exit jets. 

Valley-exit jets are a specific type of low-level jet (LLJ) \citep{Banta1996, Stensrud1996, Banta2007, Chrust2013, Tuononen2017, Nikolic2019, Svensson2019} that have been documented at canyon \citep{Banta1995, Chrust2013, Munoz2020} or valley exits \citep{Pamperin1985,Drobinski2006,Spengler2009,Sasaki2010,Hellstrom2017,Jimenez2019,Pfister2024}, where topography abruptly changes, opening onto a basin or a plain. They often arise as local modification of down-valley flows, and occur during nights with weak synoptic forcing and clear sky.
Hydraulic theory and transition of the flow from subcritical to supercritical state explain the basic dynamics: a deep layer of cold air drains out of the canyon/valley, sinks and compresses, accelerating downward into a shallower layer forming the jet, with the increase of wind speed caused by the conversion of potential into kinetic energy \citep{Chrust2013, Munoz2024}. This conversion mechanism was observed at the exit of the Inn Valley \citep{Zangl2004}, where the flow reached the maximum intensity (\SI{15}{\meter \per \second} at a height of \SI{200}{\meter} AGL) close to sunrise. Surface friction at the sidewalls may contribute to shaping the structure of the jet. The release from the frictional effects at the valley exit contributes to the acceleration of the down-valley flow \citep{Banta1995} and the approximate conservation of potential vorticity generated at the sidewalls 
leads the flow to maintain its shape beyond the exit region \citep{Zangl2004}. 

The interplay of these mechanisms shapes the valley-exit flow and its interaction with the basin air, creating a complex flow structure. Consequently, high-resolution numerical simulations are essential to investigate its propagation into the basin. The simulations performed for this work aim to answer the following research questions: does the drainage flow behave as a valley-exit wind? Is the simulation of the thermal conditions in the basin and of the valley-exit wind sensitive to the PBL scheme? How is the valley-exit wind affected by the development of a CAP in the basin? Does the valley-exit wind flow close to the ground in the basin, split or flow above the CAP?

The paper is organized as follows: Section \ref{sec: Area of interest and observations} describes the area of interest and the BTEX experiment, together with the two case studies. Section \ref{sec: Model set up} focuses on the setup and initialisation of the sub-kilometer numerical simulations. Results are presented in Section \ref{sec: Results}, where a qualitative as well as quantitative evaluation of the model performance with the different PBL schemes is provided, and in Section \ref{sec: interaction}, where the spatial structure of the drainage flow is investigated, as well as the interaction between the drainage flow and the air in the Bolzano basin. Finally, conclusions are drawn in Section \ref{sec: conclusions}.

\section{Study area and experimental measurements}
\label{sec: Area of interest and observations}

 \subsection{The Bolzano basin}
\begin{figure}[t]
\centering
\includegraphics[width=12cm]{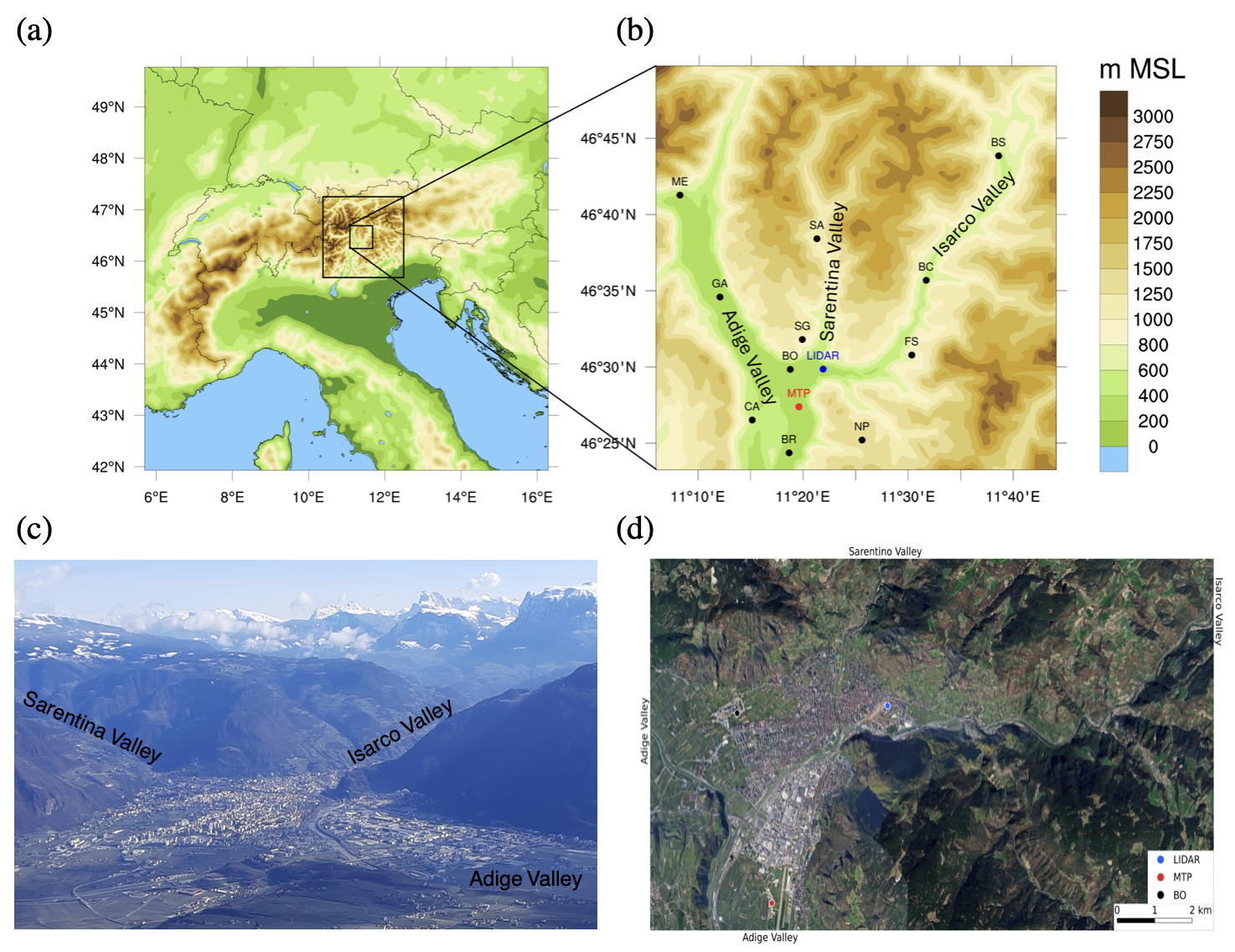}
\caption{Area of study: (a) Map of the three nested domains used for the WRF simulations, and (b) zoom in the innermost domain, with the location of the instruments used to evaluate model results, i.e. lidar, temperature profiler (MTP), and 11 ground weather stations (BR: Bronzolo, BO: Bolzano, GA: Gargazzone, ME: Merano, BC: Barbiano-Colma, CA: Caldaro, BS: Bressanone, FS: Fié allo Sciliar, SA: Sarentino, SG: San Genesio, NP: Nova Ponente). (c, d) Photo and aerial photo of the basin and the city of Bolzano with mountain crests surrounding the basin.}
\label{fig: Map domains}
\end{figure}

 The city of Bolzano (254 m a.s.l.) lies in the Adige Valley, in the northeastern Italian Alps. Here, the valley widens into a basin-like area and joins with two tributary valleys: the Sarentino Valley from North and the Isarco Valley from East. The height of mountain crests, surrounding the basin, range between \SI{1200}{\meter} and \SI{2000}{\meter} a.s.l. (Fig. \ref{fig: Map domains}). The Isarco Valley is a V-shaped valley, with a very narrow valley floor close to its exit onto the Bolzano area and a dominant southwest-northeast orientation, although it includes several bends along its length (about \SI{80}{\kilo \meter}). During nighttime, a ground-based temperature inversion frequently occurs in the Bolzano basin under weak synoptic forcing and clear sky conditions. It starts building up in the late afternoon and ends some hours after sunrise, when the energy input from solar radiation is sufficient to break up the inversion layer. The formation of a CAP in the basin, which interacts with the local circulation, is crucial for pollutant dispersion processes in the area, where significant emissions due to traffic, domestic heating and industrial activities are present \citep{Bisignano2022}, contributing to frequent poor air quality conditions.

\subsection{BTEX experiment}
The Bolzano Tracer EXperiment (BTEX) took place in the winter 2017 to study the dispersion of pollutants from a waste incinerator, located \SI{2}{\kilo\meter} southwest of the city of Bolzano \citep{Falocchi2020,Zardi2021}. During BTEX multiple instruments were used to monitor atmospheric conditions in the area, including a Doppler wind lidar, a thermal profiler and 15 ground weather stations (GWSs). The Doppler wind lidar, a WINDCUBE 100S manufactured by Leosphere (France), was installed on the roof of a building close to the exit of the Isarco Valley. It provided vertical profiles of the wind along 106 vertical levels, 10-m spaced (from \SI{335}{\meter} to \SI{1385}{\meter} a.s.l) and with a vertical resolution of \SI{25}{\meter}, averaged over 10 min. The temperature profiler, a MTP-5HE passive microwave radiometer manufactured by ATTEX (Russia), was operated by the Environmental Protection Agency of the Province of Bolzano and installed south of the city at the local airport, providing temperature measurements every 10 min and interpolated along 21 vertical levels, 50-m spaced (from \SI{250}{\meter} to \SI{1250}{\meter} a.s.l.). GWSs were operated by the Meteorological Office of the Province of Bolzano. In this work, 11 out of 15 GWSs are used, because two are located outside our domain of interest and two were not working during the selected case studies. The GWSs provide 10-minute average data and are quite homogeneously distributed over the area, covering different heights from \SI{226}{\meter} to \SI{1470}{\meter} a.s.l.. Further information on the measurement setup can be found in \citet{Falocchi2020}, while a map with their location is shown in Fig.~\ref{fig: Map domains}b. 

\begin{figure}[t]
  \includegraphics[width=\linewidth]{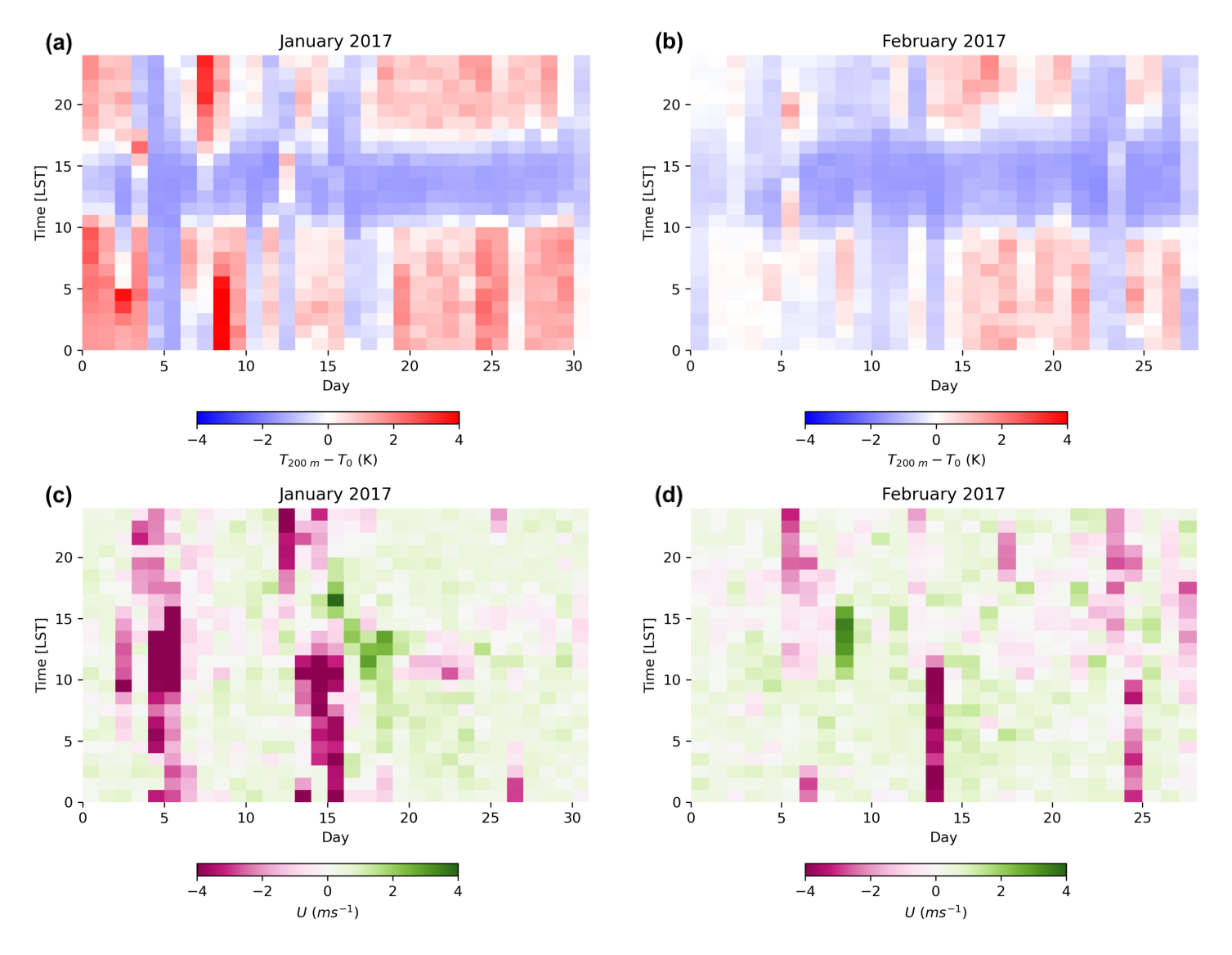}
 \caption{Monthly evolution of hourly temperature differences between \SI{450}{\meter} and \SI{250}{\meter} a.s.l. (200 m AGL and the surface) from temperature profiler measurements in (a) January and (b) February. Monthly evolution of the hourly west-east component of the 10-m wind speed measured at Bolzano GWS (see Fig. \ref{fig: Map domains}b), in (c) January and (d) February.}
\label{fig: dT vs WS bolzano}
\end{figure}

\subsection{Case studies}
\label{subsec: case studies}
In order to identify suitable case studies, an analysis of the thermal stratification in the basin, together with near-surface wind conditions, is conducted for the months of January and February 2017. 

Figure \ref{fig: dT vs WS bolzano}a,b shows the hourly temperature differences, for January and February 2017, between the measurements recorded by the temperature profiler at \SI{450}{\meter} and \SI{250}{\meter} a.s.l. (200 m AGL and the surface), with red colors corresponding to temperature inversions in the basin. While persistent CAPs are not observed, nocturnal ones frequently occur, especially in January. Comparing this situation with the west-east component of the 10-m wind speed measured at Bolzano GWS (Fig. \ref{fig: dT vs WS bolzano}c,d), it can be observed that easterly winds (pink values) during nighttime are generally registered when no temperature inversion or a weak stratification is present in the basin.

On the other hand, when a strong inversion builds up, westerly winds are generally observed during nighttime at Bolzano GWS. In February (Fig.~\ref{fig: dT vs WS bolzano}d), easterly winds are rarely observed at the ground, and the night between the 13th and 14th appears to be an exceptional case of strong easterly wind, occurring again when a thermal inversion is not present in the basin. This night and the night between the 28th and 29th January (also considering that wind lidar measurements are available from January 9th, \citealp{Falocchi2019}) are selected as case studies. These two case studies allow to analyse the interaction between the valley-exit wind of the Isarco Valley and the air in the Bolzano basin in two opposite situations, i.e., when a strong CAP is present (28-29 January, Episode 1) and without a CAP (13-14 February, Episode 2). 

In the 28-29 January 2017 case, northern Italy was under a high-pressure system with no appreciable synoptic winds and the area of Bolzano was cloud-free. 

The 13-14 February 2017 episode was similarly characterised by high-pressure conditions over northern Italy, but in this case weak southeasterly synoptic winds transported moisture from the Mediterranean towards the Alps, as evidenced by an increase in absolute humidity recorded by many GWSs at varying altitudes (see Fig.~6c in \citeauthor{Falocchi2020}, \citeyear{Falocchi2020}). As a consequence, stratocumulus clouds were present above the Adige Valley during the night, hindering radiative cooling and the consequent formation of a ground-based temperature inversion in the basin. Satellite images show that clouds were restricted to the Adige Valley and the Bolzano basin, leaving the Isarco Valley cloud-free and leading to a strong thermal gradient between the tributary valley, with cooler air, and the basin, with the development of a nocturnal down-valley flow. 

\section{Model setup}
\label{sec: Model set up}
Simulations are performed with the WRF model version 4.3.3 \citep{Skamarock2019}. Each simulation lasts 42 hours and starts the day before the night of interest (28 January 00 UTC for Episode 1 and 13 February 00 UTC for Episode 2). The first 14 hours are considered spin-up time. Three two-way nested grids are used (Fig. \ref{fig: Map domains}a) with $196^2$, $196^2$, $166^2$ horizontal cells ($\Delta x = $\SI{4.5}{\kilo \meter}, \SI{0.9}{\kilo \meter}, \SI{0.3}{\kilo \meter}), while 61 levels are used in the vertical direction, with 24 levels below \SI{1000}{\meter} AGL and the first model level at \SI{14}{\meter} AGL. The inner domain covers the area of interest (Fig. \ref{fig: Map domains}b).

Four simulations for each episode are performed, with the following choices of PBL scheme:  1) the Yonsei State University (YSU, \citealp{Hong2006YSU}), 2) the Mellor-Yamada-Janjic (MYJ, \citealp{Janjic2002}), 3) the Bougeault-Lacarr\'{e}re (BouLac, \citealp{Bougeault1989}) and 4) a k-$\epsilon$-based closure recently proposed in \cite{Zonato2022} (KEPS-TPE).
The YSU PBL scheme is a first-order closure, while MYJ and BouLac are 1.5-order closures with a prognostic equation for the TKE, with MYJ based on \citet{Mellor1982} (level 2.5) and BouLac on \citet{Therry1983}. The KEPS-TPE PBL scheme is a k-$\epsilon$-based closure, including prognostic equations for the TKE, its dissipation rate and the temperature variance. The latter is directly connected to the TPE. The heat flux parameterization includes a counter-gradient term proportional to the temperature variance and accounts for the self-control mechanism of stratified turbulence \citep{Zilitinkevich2007}. YSU, BouLac and KEPS-TPE are coupled to the MM5 surface layer scheme \citep{Grell1994} and MYJ to the MYJ Eta scheme \citep{Janjic1994} to calculate surface-atmosphere exchange.

The microphysics scheme adopted is WSM6 \citep{Hong2006micro}, while the RRTM scheme \citep{Mlawer1997} is used for long-wave radiation and the \citet{Dudhia1989} scheme for short-wave radiation, including the effects of slope inclination and topographic shading. No cumulus parameterization is used in any of the domains but the microphysics scheme was activated to reproduce the stratocumulus clouds, observed in the valley in Episode 2. The land surface model (LSM) adopted is Noah-MP \citep{Niu2011}.

Meteorological data from the ERA5 reanalysis released by the European Centre for Medium-Range Weather Forecasts (ECMWF) are used as initialisation and lateral boundary conditions every 6 hours, with a horizontal resolution of $\sim$\SI{30}{\kilo \meter} on 137 vertical atmospheric levels and at the surface. Several preliminary simulations were performed in order to improve the initial surface conditions. It is well known that increasing the spatial resolution of simulations without incorporating improved surface data does not automatically imply an improvement of the model performance \citep{Chow2006}. In particular, attention was paid to soil moisture and snow cover that influence the development of thermally-driven circulations and the near-surface stratification \citep{Ookouchi1984,BantaGannon1995,Massey2016,Tomasi2017,Jimenez2019}. 

ERA5 overestimates snow cover for both case studies, with snow cover even inside the Adige Valley, which was not present. On the basis of photographs taken during the field campaign, the snow cover was limited to heights above \SI{2000}{\meter} AGL. For consistency with this modification, the land-surface process that allows for the presence of supercooled liquid water in the soil when the soil temperature is at or below the freezing point, such as when snow is present, was forced to be off. Compared to the default configuration, this modification gave a more realistic distribution of the soil temperature, especially in valleys, and lower soil temperatures above \SI{2000}{\meter} AGL. At those heights, a deep snow layer isolates the soil from the atmosphere above, thus the lower soil temperature are expected not to impact the near-surface air temperature at those heights. Concerning soil moisture, unfortunately no in situ observations were available to improve model initialisation. Therefore, observations from NASA's Soil Moisture Active Passive mission (SMAP, \url{https://worldview.earthdata.nasa.gov}), with a horizontal resolution of \SI{9}{\kilo \meter}, were adopted to obtain representative values of soil moisture in the area. For Episode 1, the comparison between SMAP soil moisture values and model initialisation suggested halving ERA5 soil moisture. A similar overestimation was found in \citet{Giovannini2014} in the Adige Valley. For Episode 2, instead, no adjustment was needed. 

High-resolution topography data were obtained from the Viewfinder Panoramas website (http://www.viewfinderpanoramas.org), with a spatial resolution of 1 arc-s ($\sim$\SI{30}{\meter}), and slightly smoothed to prevent numerical instability. Similarly, the Corine Land Cover (CLC) dataset updated to 2012, provided by the European Environment Agency (http://www.eea.europa.eu), with a spatial resolution of \SI{100}{\meter}, was employed as land use dataset after a proper reclassification of CLC classes into the 20 MODIS classes adopted by WRF, following \citet{Giovannini2014}. 

\section{PBL schemes intercomparison}
\label{sec: Results}
\subsection{28-29 January 2017: Episode 1}
\label{sec: results2829}
\begin{figure}[p!]
\centering
\includegraphics[width=0.7\linewidth]{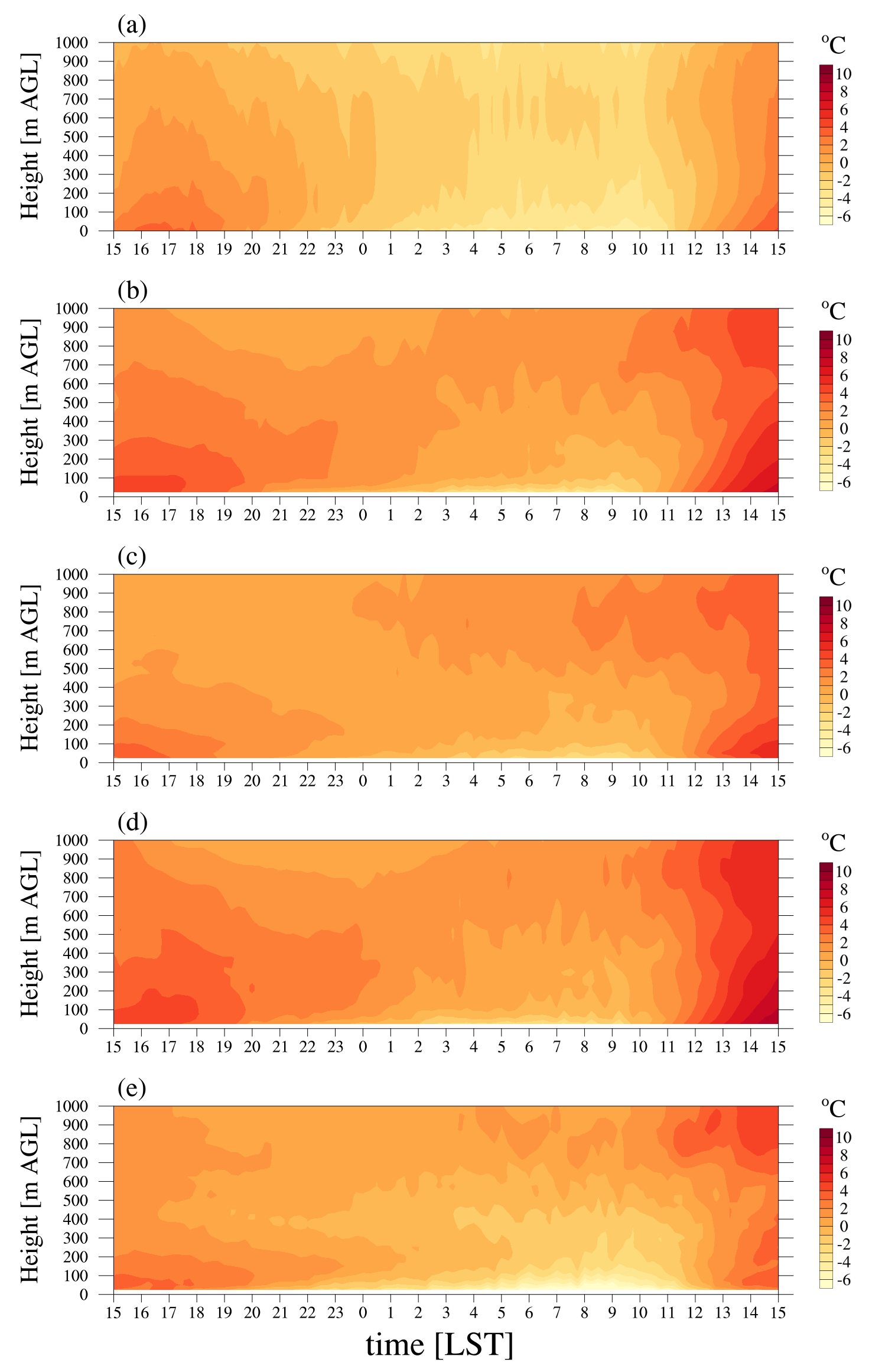}
\caption{Time-height plots of temperature for Episode 1 at the temperature profiler site as (a) measured and simulated by (b) YSU, (c) MYJ, (d) BouLac and (e) KEPS-TPE.}
\label{fig: Fig4_profiler}
\end{figure}

\begin{figure}[p!]
\centering
\includegraphics[width=0.7\linewidth]{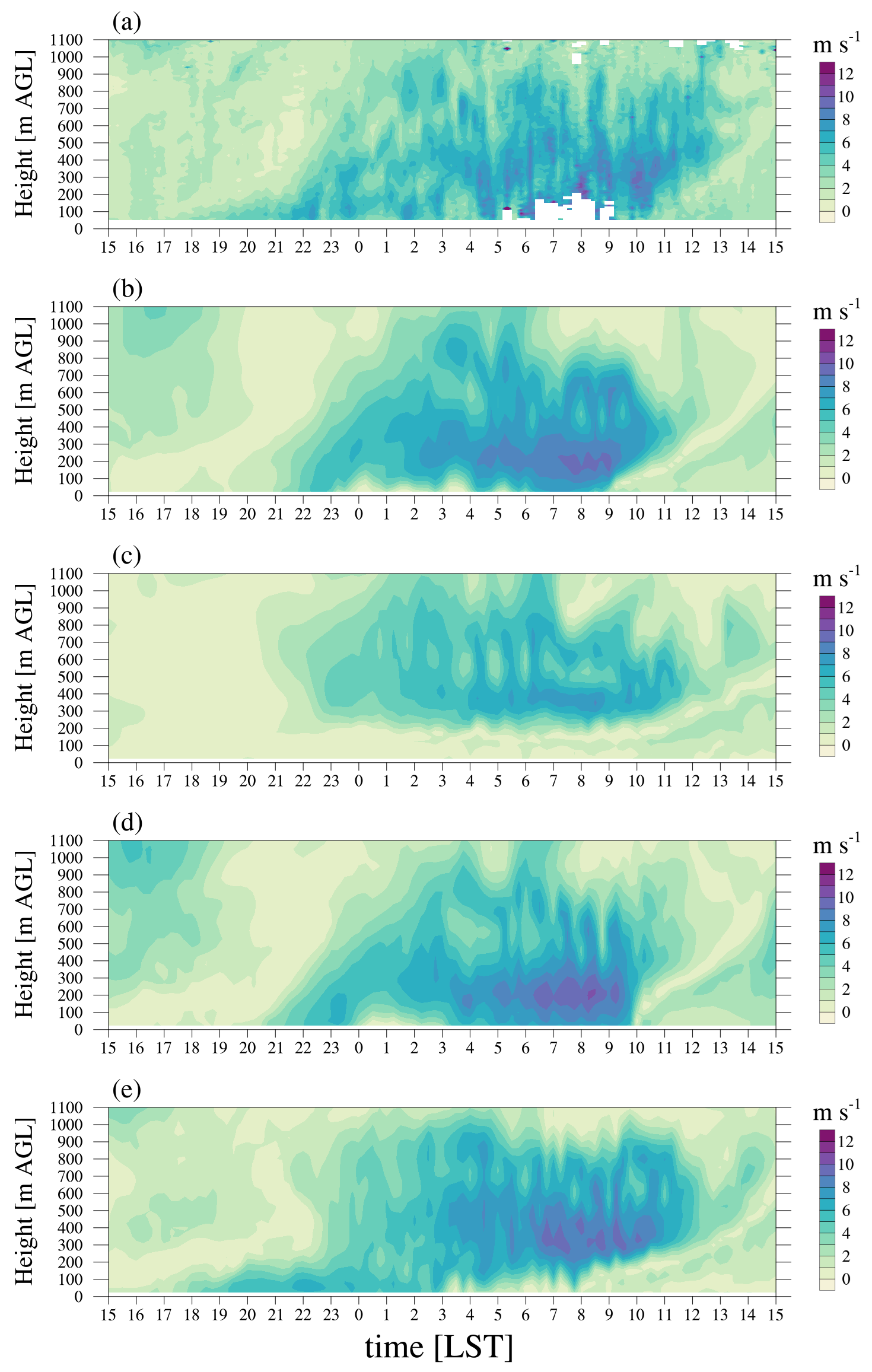}
\caption{Time-height plots of horizontal wind speed for Episode 1 at the lidar site as (a) measured and simulated by (b) YSU, (c) MYJ, (d) BouLac and (e) KEPS-TPE}
\label{fig: Fig5_lidar}
\end{figure}
\begin{figure}[p!]
\centering
\includegraphics[width=0.7\linewidth]{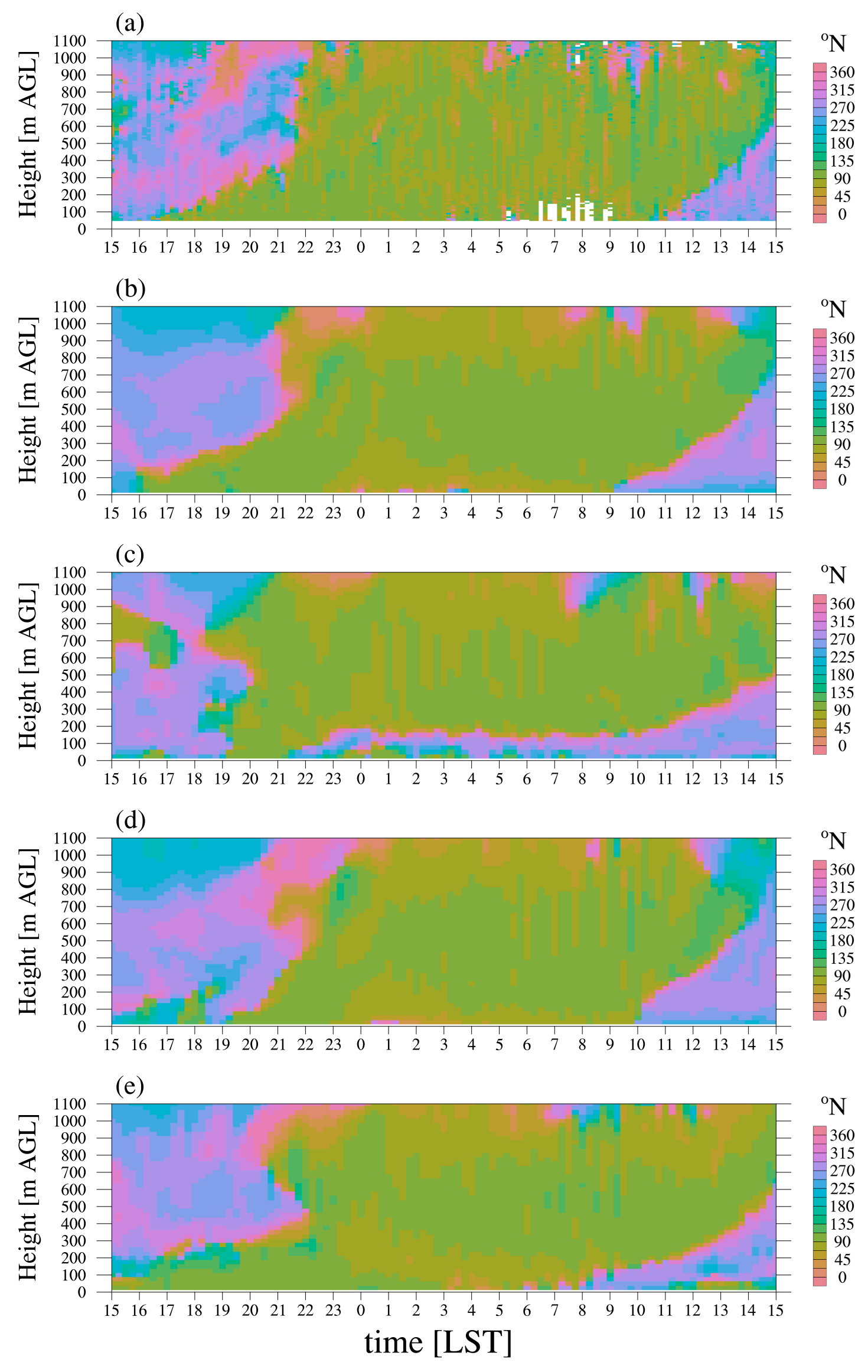}
\caption{Time-height plots of horizontal wind direction for Episode 1 at the lidar site as (a) measured and simulated by (b) YSU, (c) MYJ, (d) BouLac and (e) KEPS-TPE.}
\label{fig: Fig6_lidar_dir}
\end{figure}
Figure~\ref{fig: Fig4_profiler} shows the time-height plots of temperature from the thermal profiler and the simulations for Episode 1. An inversion layer started to build up a few hours after sunset (local sunset around 16:00 LST), reaching a top height of \SI{700}{\meter} AGL during the night, with a strong stratification close to the surface. The ground-based temperature inversion persisted for some time after sunrise (local sunrise around 10:00 LST) before being eroded from the lowest atmospheric layers. 

While none of the PBL schemes is able to reproduce the depth of the CAP, the KEPS-TPE PBL scheme aligns most closely with observations, yielding the coldest near-surface temperatures. This behaviour is attributed to the counter-gradient term included in the scheme. By counteracting the purely local (down-gradient) diffusive component, this term weakens the coupling between the lowest layers and the atmosphere aloft. Consequently, this reduces downward heat transport, specifically the replenishment of warmer air from the residual layer toward the surface layer.

The time-height plots of the observed wind speed and direction at the lidar location (Figs~\ref{fig: Fig5_lidar}a and \ref{fig: Fig6_lidar_dir}a) highlight that the drainage (easterly) flow started to develop around 19:00 LST after the formation of the thermally stable layer near the surface. The wind speed became continuously stronger over an increasingly deep layer up to a height of \SI{900}{\meter} AGL, following the evolution of the thermal structure into the basin. Peaks in wind speed were reached in the morning at 08:00 LST below \SI{400}{\meter} AGL, with an intensity of $\sim$\SI{10}{\meter \per \second}. Then, the drainage flow rapidly died out, starting from the surface. All PBL schemes reproduce the onset of the drainage wind exiting the Isarco Valley with a slight delay, although KEPS-TPE shows the best agreement with observations. All schemes, except MYJ, reproduce well the structure of the valley-exit wind, whose evolution is strictly connected with the growth of the ground-based temperature inversion. However, MYJ tends to overestimate the base of the drainage flow, reproduced at \SI{200}{\meter} AGL. 

\begin{figure}[h!]
\centering
\includegraphics[width=0.7\linewidth]{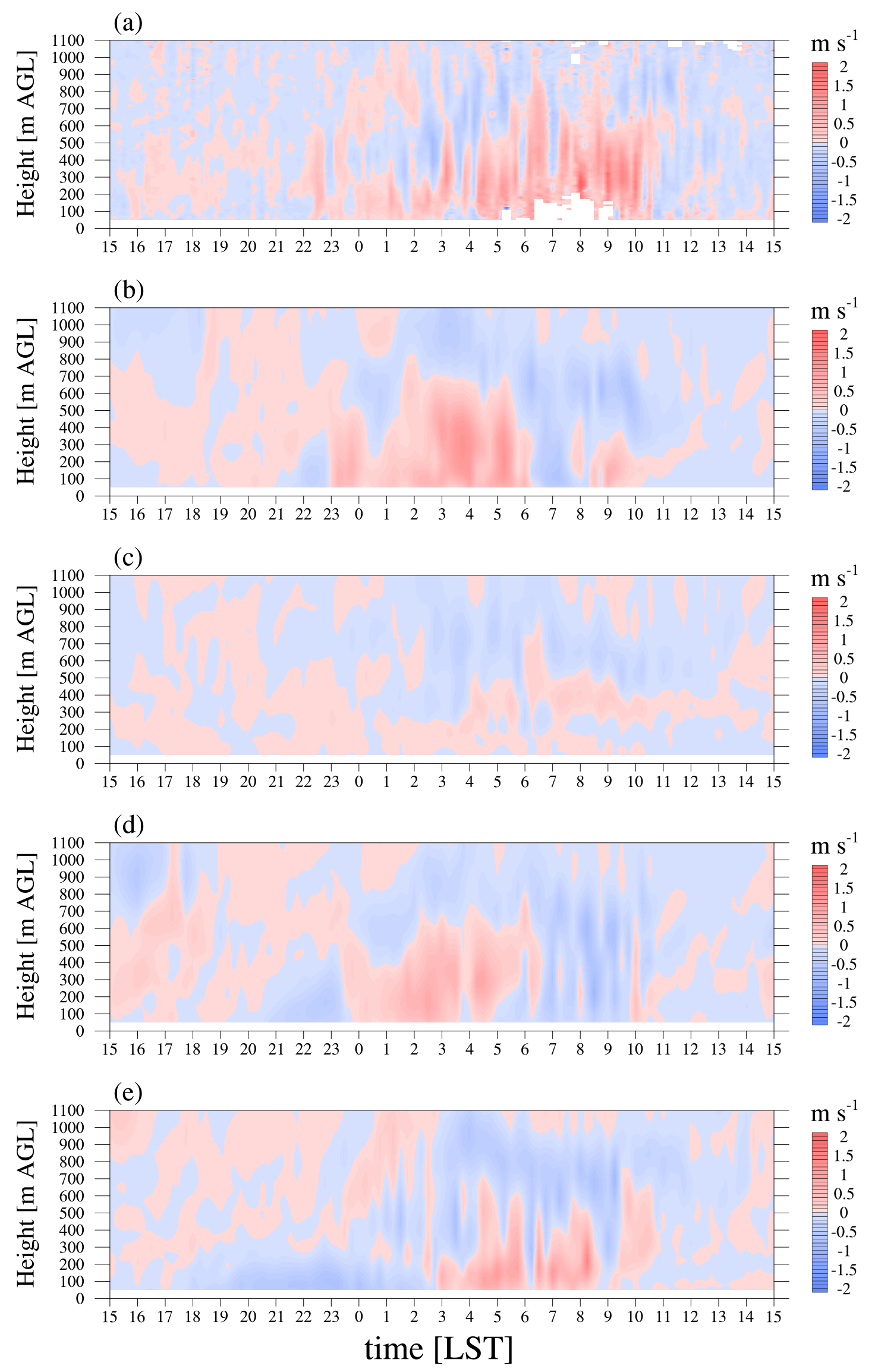}
\caption{Time-height plots of vertical wind speed for Episode 1 at the lidar site as (a) measured and simulated by (b) YSU, (c) MYJ, (d) BouLac and (e) KEPS-TPE.}
\label{fig: Fig7_lidar_w}
\end{figure}

The lidar measurements of the vertical wind component are shown in Fig.~\ref{fig: Fig7_lidar_w}a. 
The valley-exit wind is associated with positive vertical velocities at the lidar location, strengthening with the increase in the horizontal wind speed in the second part of the night and in the early morning (Fig. \ref{fig: Fig5_lidar}a). The simulations capture the upward vertical velocity, except for MYJ, which reproduces a vertical wind speed much weaker than observations. In the MYJ simulation, the CAP is more extended towards the Isarco Valley compared to the other simulations, with a strong upward vertical velocity observed a few cells further east. This explains why the base of the valley-exit wind is reproduced at higher heights (Figs.~\ref{fig: Fig5_lidar}c and \ref{fig: Fig6_lidar_dir}c). The upward motion in KEPS-TPE starts later, around 03:00 LST, and, again, the mismatch is related to the particular grid cell used for this analysis: a better agreement with observations would be obtained with cells immediately east of the lidar location. 

\subsubsection{Quantitative evaluation of model results}
For a quantitative overview of the model performance with the different PBL schemes, results and observations are compared in terms of mean error (ME) and root-mean-square error (RMSE). The atmospheric conditions at the surface are evaluated against measurements from the GWSs. ME and RMSE are estimated from hourly averages in the period 28-01-2017 15:00 LST to 29-01-2017 18:00 LST, and are summarized in Table~\ref{tab: GWS_T_2829} and Table~\ref{tab: GWS_WS_2829} for 2-m temperature and 10-m wind speed, respectively, with minimum ME and RMSE values at each GWS in bold. 

For 2-m temperature, the ME is generally positive for all the simulations, meaning an average model overestimation, with the exception of KEPS-TPE, which shows underestimations at most GWSs. 
Overall, the best performance is attained by YSU, as shown also by the RMSE values, and by the errors averaged over the ensemble of GWSs (<ME>, <RMSE>). 
\begin{table}[bt]
\caption{ME and RMSE for the 2-m temperature of the different simulations for Episode 1. ME and RMSE values are computed from hourly averages of GWS measurements and WRF outputs in the interval 28-01-2017 15:00 LST - 29-01-2017 18:00 LST. <ME> and <RMSE> refer to the average values of the metrics considering all the GWSs.}
\label{tab: GWS_T_2829}
\begin{tabular}{lllllllrrrc}  
\headrow
\thead{\#} & \thead{GWS} & \thead{Height (m a.s.l)} & \thead{} & \thead{} & \thead{ME} & \thead{}& \thead{} & \thead{} & \thead{RMSE} & \thead{}\\
\thead{T2} & \thead{} & \thead{} & \thead{YSU} & \thead{MYJ} & \thead{BouLac} & \thead{KEPS-TPE}& \thead{YSU} & \thead{MYJ} & \thead{BouLac} & \thead{KEPS-TPE}\\
\hiderowcolors
1 & BR & 226 & \textbf{0.1}  & 0.6 & 1.4 & -2.4 &  \textbf{1.5} & 2.4 &  2.0&3.4\\
2 & B0 & 254 & \textbf{-0.2}  & 0.8 & 1.1 & -2.3 &  \textbf{1.2} & 2.3 &  1.8&3.1\\
3 & GA & 254  & 1.5  & 2.7 & 3.0 & \textbf{-1.2} &  \textbf{2.7} & 4.1 & 3.9 & 3.3\\
4 & ME & 330  & 2.3  & 2.0 & 3.0 & \textbf{1.0} &  3.0 & \textbf{2.7} &  3.5&  3.0\\
5 & BC & 490 & 2.3   & 2.4 & 3.0 & \textbf{-0.8} &  2.9 & 3.1 & 3.6& \textbf{1.5}\\
6 & CA & 495 & \textbf{-0.1}  & -0.8 & 0.8 & -1.9 & \textbf{0.9}  & 1.3 & 1.2 &2.1\\
7 & BS & 585 & 3.1   & 3.3 &  3.9 & \textbf{0.9} &  3.5  & 3.9 & 4.2 & \textbf{2.4}\\
8 & FS & 840 &  \textbf{0.3}  & 0.4 & 1.1 & -0.5 &  \textbf{0.9} & 1.4 &1.2  &1.4\\
9 & SA & 970 & 1.8   & 2.0 & 2.3 & -1.8 &  \textbf{2.0} & 2.5 & 2.4 &2.9\\
10 & SG & 970 & 2.7   & 2.6 & 3.2 & \textbf{2.2} &  3.1 & 3.1 & 3.4&\textbf{3.0}\\
11 & NP & 1470 & -0.6 & -0.9 & \textbf{-0.4} & -1.1 &  1.5 & 1.7 & \textbf{1.4}& 1.6\\
\hline  
& & \textbf{<ME> $\quad$ <RMSE>} & \textbf{1.2}	&1.4	&2.0	&-0.7	&\textbf{2.1}&	2.6	&2.6 & 2.5\\

\end{tabular}
\end{table}

\begin{table}[bt]
\caption{ME and RMSE for the 10-m wind speed of the different simulations for Episode 1. ME and RMSE values are computed from hourly averages of GWS measurements and WRF outputs in the interval 28-01-2017 15:00 LST - 29-01-2017 18:00 LST. The symbol \textsuperscript{*} means $|\textrm{ME}|$ less than 0.05. <ME> and <RMSE> refer to the average values of the metrics considering all the GWSs.}
\label{tab: GWS_WS_2829}
\begin{tabular}{lllllllrrrc}  
\headrow
\thead{\#} & \thead{GWS} & \thead{Height  (m a.s.l)} & \thead{} & \thead{} & \thead{ME} & \thead{}& \thead{} & \thead{} & \thead{RMSE} & \thead{}\\
\thead{WS10} & \thead{} & \thead{} & \thead{YSU} & \thead{MYJ} & \thead{BouLac} & \thead{KEPS-TPE}& \thead{YSU} & \thead{MYJ} & \thead{BouLac} & \thead{KEPS-TPE}\\
\hiderowcolors
1 & BR & 226 & -0.5  & \textbf{0.0\textsuperscript{*}} &-0.3& -0.3 &  0.7 & 0.8& 0.7&0.7\\
2 & BO & 254 & -0.1  & -0.1 & \textbf{0.0\textsuperscript{*}}& -0.1&  0.4 & \textbf{0.3}& 0.5 &0.4\\
3 & GA & 254  & 0.3  & \textbf{0.1} & 0.3& 0.2 &  0.6 & 0.6& 0.6& 0.6\\
4 & ME & 330  & 0.1  & 0.1 & \textbf{0.0\textsuperscript{*}} & 0.2 &  0.5 & 0.6 &   0.5&  \textbf{0.4}\\
5 & BC & 490 & 2.0   & 0.5 & 1.7 & \textbf{0.3} &  2.3 & 1.2&2.0& \textbf{0.8}\\
6 & CA & 495 & 0.2  & -0.8 & 0.1& -0.1 & 0.7  & 1.3 & 0.7 &0.9\\
7 & BS & 585 & 0.3   & -0.1 & 0.4 & \textbf{0.0\textsuperscript{*}} &  0.7 & \textbf{0.3}&0.7& 0.4\\
8 & FS & 840 &  -0.1  & \textbf{0.0\textsuperscript{*}}& 0.1 & -0.1 &  0.3 & 0.3& 0.3 &0.5\\
9 & SA & 970 & -0.2   & -0.4 & \textbf{0.0\textsuperscript{*}} & 0.9 &  \textbf{0.4} & 0.5 & 0.5&1.7\\
10 & SG & 970 & 0.3   & \textbf{0.0\textsuperscript{*}} & 0.2& 0.2&  0.6 & 0.4& 0.4&0.6\\
11 & NP & 1470 & -0.2 & -0.7& -0.2&-0.4 &  0.6 & 0.9& 0.6& 0.7\\
& & \textbf{<ME> $\quad$ <RMSE>}& 0.2 & -0.1 & 0.2 & 0.1 & 0.7 & 0.7 & 0.7 & 0.7 \\
\hline  
\end{tabular}
\end{table}
Across all simulations, ME and RMSE values for 10-m wind speed are consistently low, generally remaining below \SI{1}{\meter \per \second} regardless of GWS location. Additionally, the spatially averaged metrics are comparable across the different simulations.

\begin{figure}[h!]
\centering
\includegraphics[width=\linewidth]{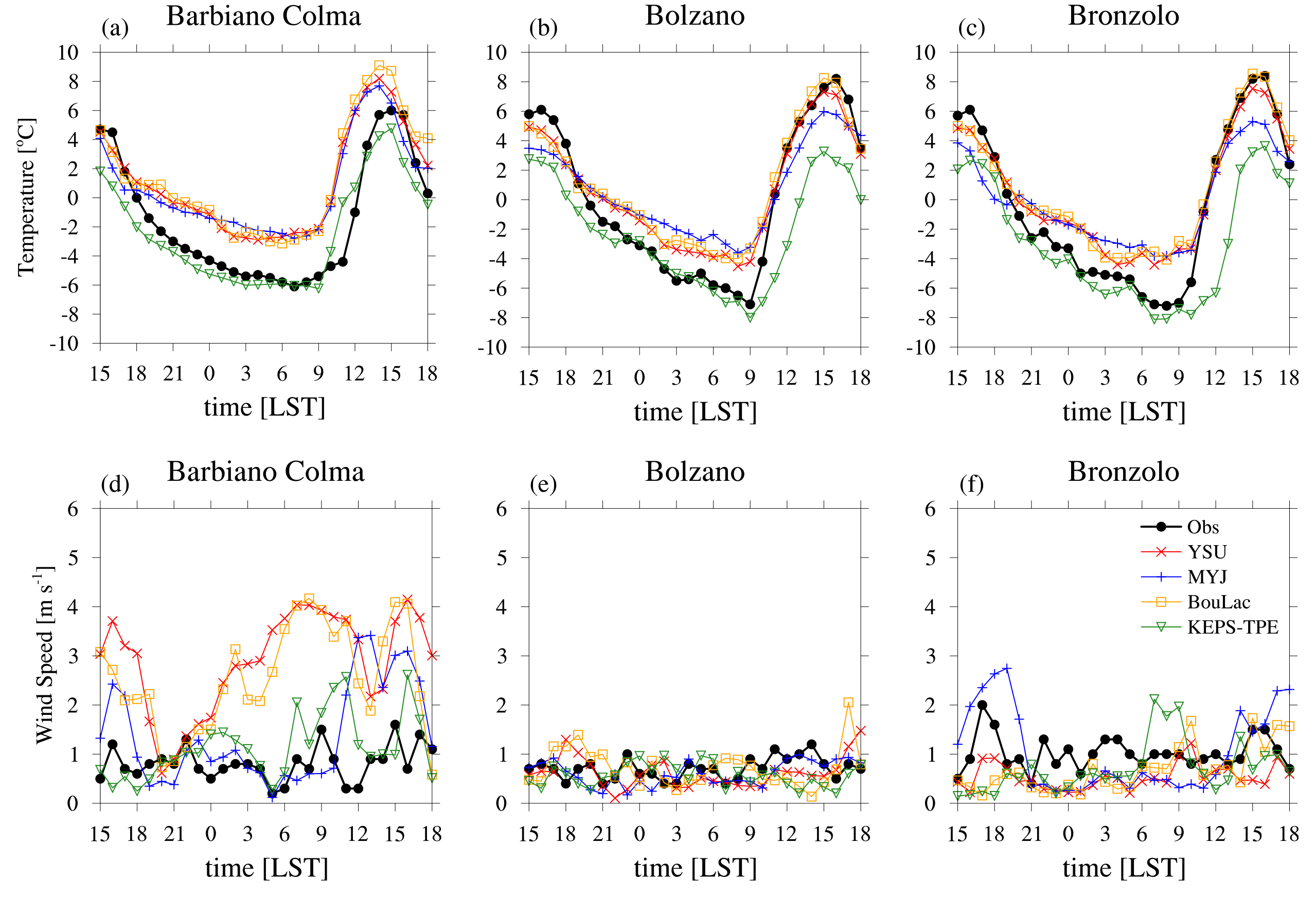}
\caption{Hourly time series of 2-m temperature (upper panel) and 10-m wind speed (lower panel) at the GWSs of Barbiano-Colma, Bolzano and Bronzolo from observations (dotted black line) and simulations for Episode 1.}
\label{fig: GWS_2829_BBB}
\end{figure}

In order to understand more in detail the performance of the PBL schemes in reproducing 2-m temperature and 10-m wind speed, the time series of three GWSs, representative of the Isarco Valley (Barbiano-Colma), the Bolzano basin (Bolzano) and the Adige Valley (Bronzolo), are shown in Fig.~\ref{fig: GWS_2829_BBB}. All PBL schemes overestimate temperatures during nighttime, except KEPS-TPE, which is in good agreement with observations. However, during daytime, KEPS-TPE tends to significantly underestimate temperatures, especially at Bolzano and Bronzolo. This suggests that the overall ME for this scheme, shown in Table~\ref{tab: GWS_T_2829}, should be attributed to daytime conditions. During daytime, temperatures are slightly underestimated also by MYJ at Bolzano and Bronzolo, whereas BouLac displays a very good agreement with observations. All schemes, except KEPS-TPE, tend to overestimate the near-surface temperature at Barbiano-Colma, i.e., in the Isarco Valley, during daytime.

All PBL schemes reproduce a weak and mostly constant wind speed at 10-m AGL at Bolzano and Bronzolo, in agreement with observations. During nighttime YSU and BouLac overestimate wind speed at Barbiano-Colma, likely associated with the onset of the drainage flow in the Isarco Valley. Thus, these two schemes reproduce a drainage flow close to the ground, which is not confirmed by observations. Model errors at Barbiano-Colma may be due to the complex topography of the valley, which is very narrow at this location. Local atmospheric conditions, such as a ground-based temperature inversion interacting with the drainage flow, may not be correctly captured because the horizontal resolution of the model is not sufficient to resolve all the fine topographic details influencing this interaction.

\begin{figure}[p!]
\centering
\includegraphics[width=0.7\linewidth]{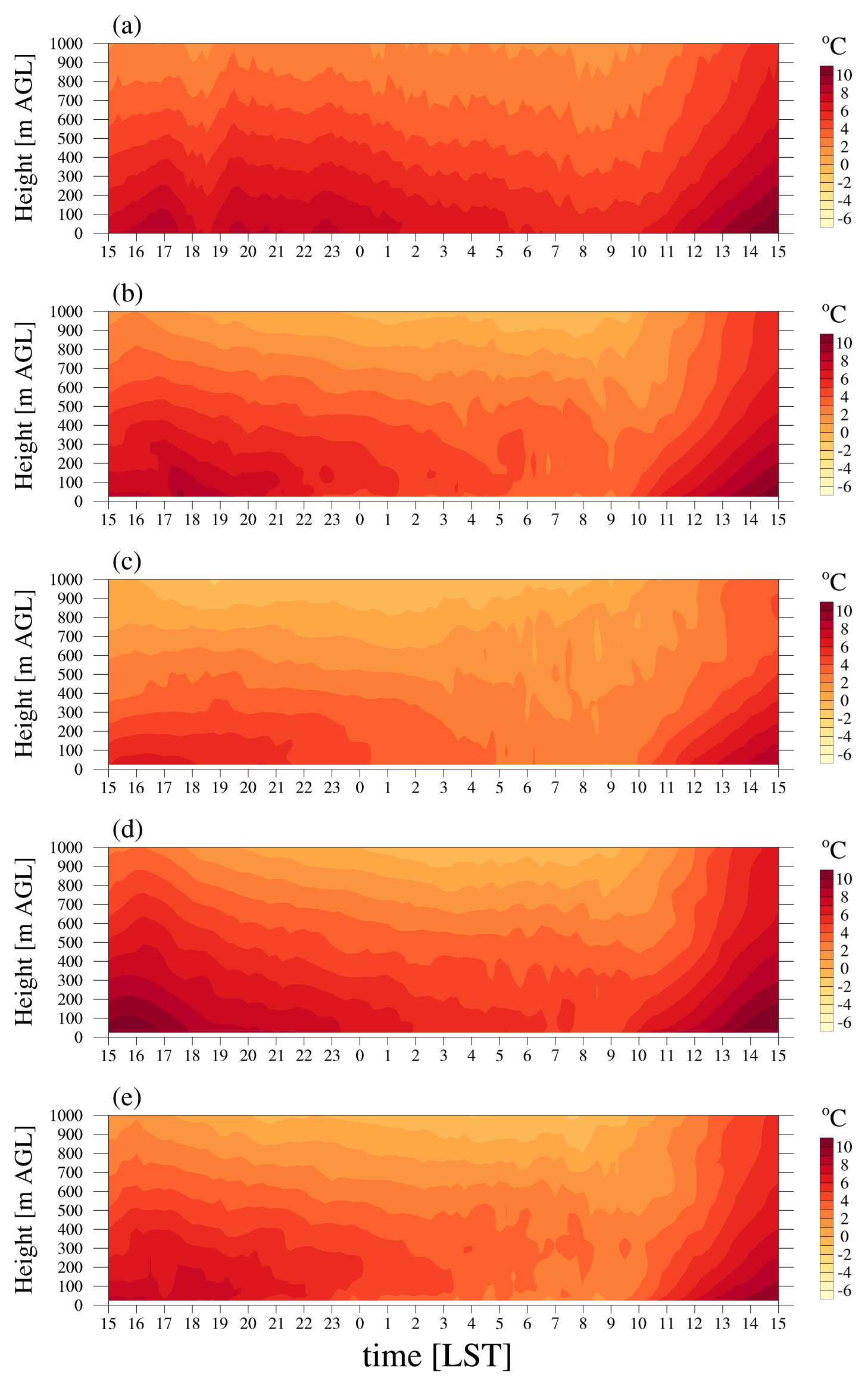}
\caption{Time-height plots of temperature for Episode 2 at the temperature profiler site as (a) measured and simulated by (b) YSU, (c) MYJ, (d) BouLac and (e) KEPS-TPE.}
\label{fig: Fig9_1314_profiler}
\end{figure}
\begin{figure}[p!]
\centering
\includegraphics[width=0.7\linewidth]{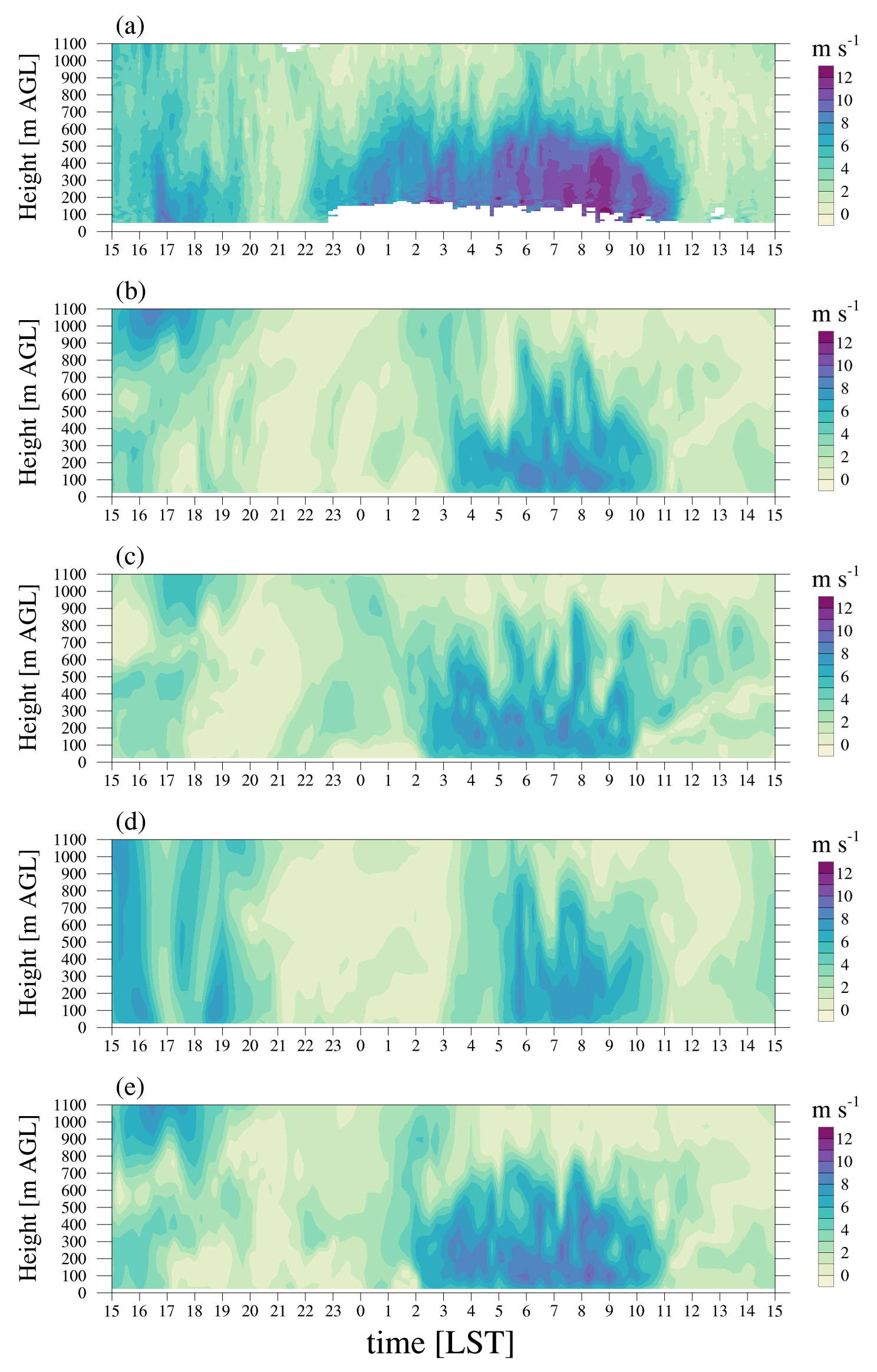}
\caption{Time-height plots of horizontal wind speed for Episode 2 at the lidar site as (a) measured and simulated by (b) YSU, (c) MYJ, (d) BouLac and (e) KEPS-TPE.}
\label{fig: Fig10_1314_lidar}
\end{figure}

\begin{figure}[p!]
\centering
\includegraphics[width=0.7\linewidth]{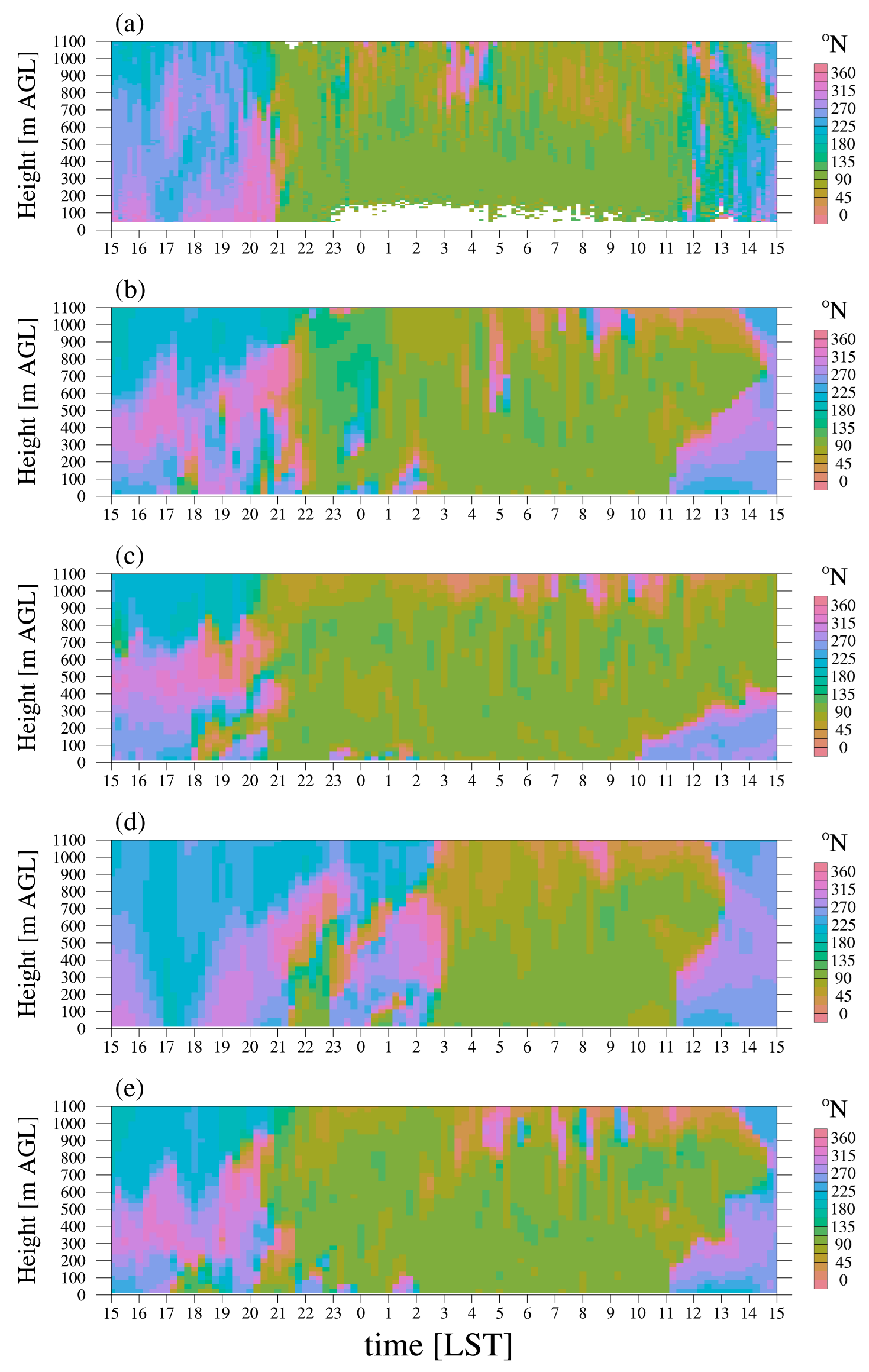}
\caption{Time-height plots of horizontal wind direction for Episode 2 at the lidar site as (a) measured and simulated by (b) YSU, (c) MYJ, (d) BouLac and (e) KEPS-TPE.}
\label{fig: Fig11_1314_lidar_dir}
\end{figure}

\subsection{13-14 February 2017: Episode 2}
\label{sec: Results_ep2}
Figure~\ref{fig: Fig9_1314_profiler} shows the temperature time-height plots from the thermal profiler and the simulations for Episode 2. The temperature in the basin started to decrease after sunset, uniformly across all the atmospheric layers investigated. A CAP did not form in this case, due to the presence of stratocumulus clouds over the Adige Valley. During daytime, all simulations show a good agreement with observations except for MYJ, which underestimates temperature, particularly on the first day. A comparison between the surface sensible heat flux from the different simulations did not show significant differences, thus excluding the different surface-layer scheme coupled with MYJ (Sec. \ref{sec: Model set up}) as a possible cause of this underestimation (not shown). 
During the night, all schemes reproduce quite well the vertical structure of the atmosphere, but they mostly overestimate the cooling after sunset, especially in the evening and in the early night. This results in a temperature underestimation of \SI{2}-\SI{3} {\degreeCelsius} along the whole vertical layer investigated, except for BouLac. The latter simulates a nocturnal cooling comparable to the other schemes, but, being characterized by higher temperature during daytime, it displays warmer temperatures during the night and in the following morning.

The time-height plots of wind speed and direction at the lidar location from observations (Figs~\ref{fig: Fig10_1314_lidar}a and \ref{fig: Fig11_1314_lidar_dir}a) show a concurrent increase of the wind speed at different heights around 22:00 LST, up to \SI{600}{\meter} AGL, with a sharp change from westerly to easterly winds. The onset of the drainage flow was more uniform compared to the gradual deepening observed in Episode 1, and the increase in wind intensity was more rapid throughout the night, reaching a peak of $\sim$\SI{12}{\meter \per \second} at 09:00 LST. The drainage flow rapidly died out a few hours after sunrise, around 11:00 LST. All PBL schemes reproduce the onset with a delay, more significant in BouLac and YSU. A significant underestimation of wind speed is observed throughout the night, with KEPS-TPE being the scheme closer to observations. YSU and KEPS-TPE reproduce better the intensification of the drainage flow and the velocity peak before sunrise, which is well reproduced especially by KEPS-TPE with a maximum exactly between 08:00 LST and 09:00 LST. In the early afternoon of the first day, all schemes agree on a north-westerly wind (Fig.~\ref{fig: Fig11_1314_lidar_dir}), probably associated with a southerly synoptic wind channeling into the Adige Valley \citep[a mechanism described in][]{Whiteman1993} and presenting as an up-valley wind in the Isarco Valley.
This channeling mechanism strengthens the up-valley wind, more intense in this case compared to Episode 1. This effect is simulated quite strongly by BouLac, which is also the scheme exhibiting the largest delay in the onset of the valley-exit wind, suggesting that the stronger up-valley flow impedes the development of the nocturnal local circulation.  

\begin{figure}[h!]
\centering
\includegraphics[width=0.7\linewidth]{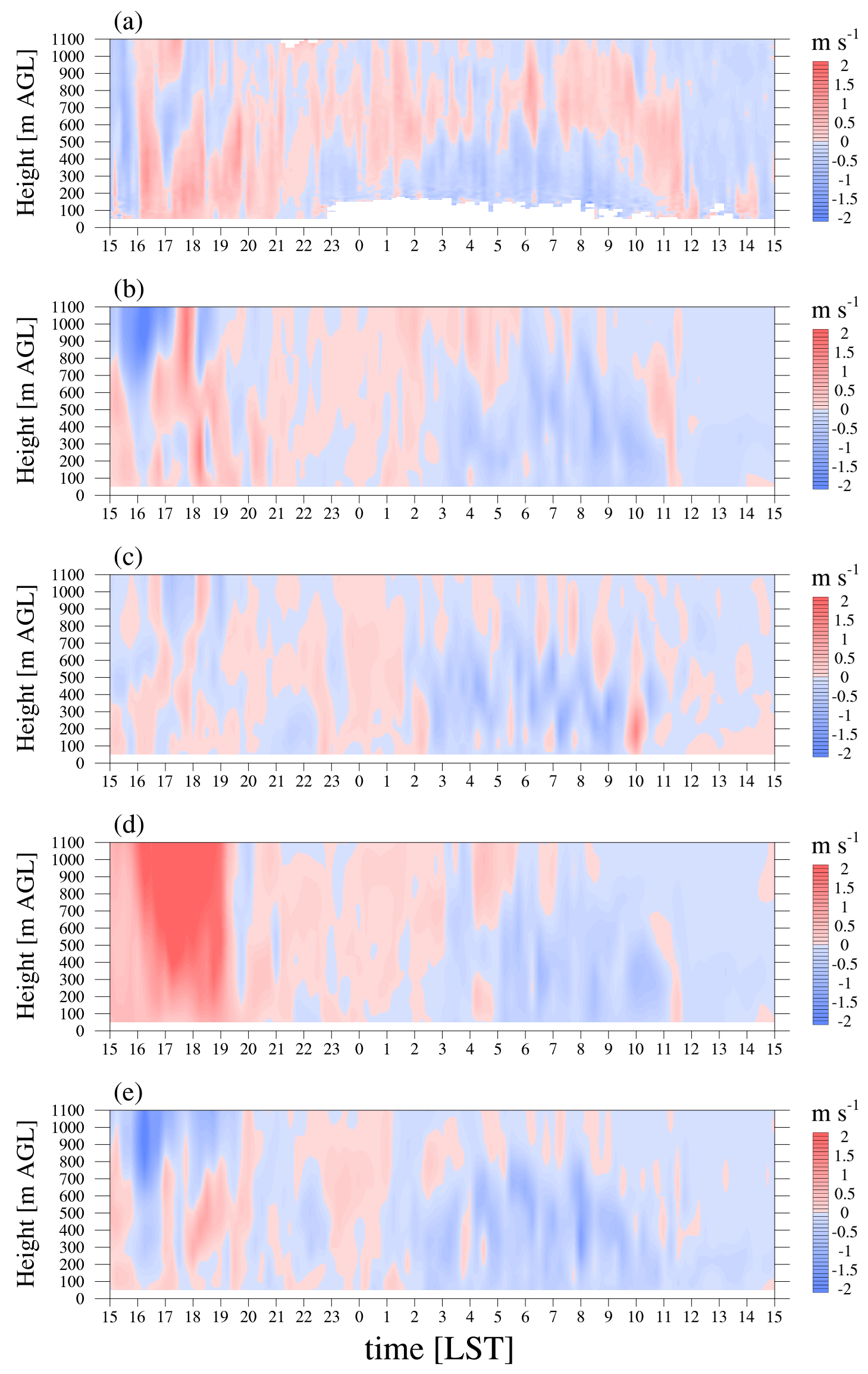}
\caption{Time-height plots of vertical wind speed for Episode 2 at the lidar site as (a) measured and simulated by (a) YSU, (b) MYJ, (c) BouLac and (d) KEPS-TPE.}
\label{fig: Fig12_1314_lidar_w}
\end{figure}

Figure~\ref{fig: Fig12_1314_lidar_w} shows the vertical wind speed from the lidar and the simulations. A downward flow component (negative values) was observed upon the development of the valley-exit wind. The model accurately reproduces this feature across all PBL schemes, with temporal shifts corresponding to the specific onset of the valley-exit wind in each simulation.
\begin{figure}[bt]
\centering
\includegraphics[width=\linewidth]{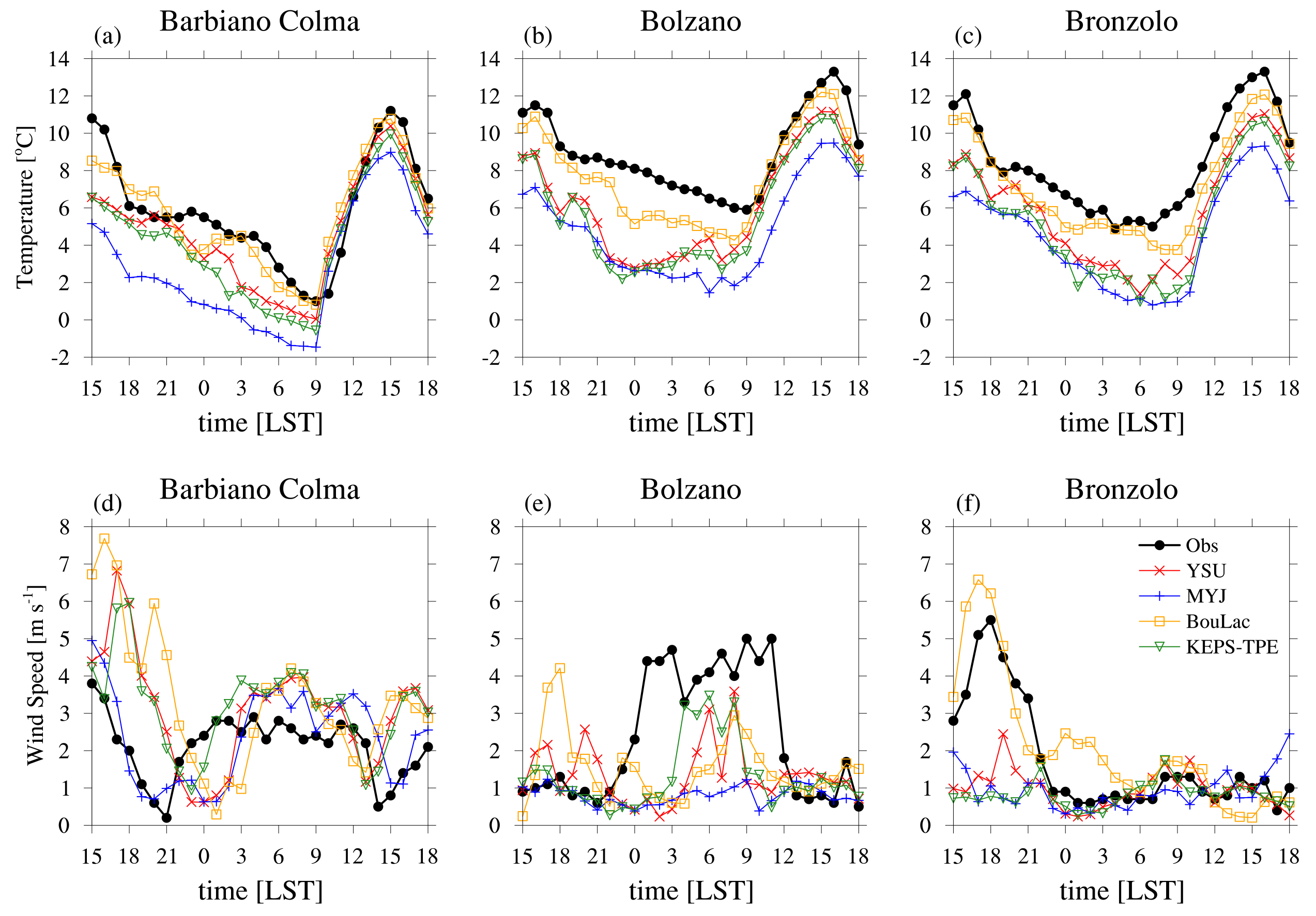}
\caption{Hourly time series of 2-m temperature (upper panel) and 10-m wind speed (lower panel) at the GWSs of Barbiano-Colma, Bolzano and Bronzolo from observations (dotted black line) and simulations for Episode 2.}
\label{fig: GWS_1314_BBB}
\end{figure}

\begin{table}[bt]
\caption{ME and RMSE for the 2-m temperature of the different simulations for Episode 2. ME and RMSE values are computed from hourly averages of GWS measurements and WRF outputs in the interval 13-02-2017 15:00 LST - 14-02-2017 18:00 LST. <ME> and <RMSE> refer to the average values of the metrics considering all the GWSs.}
\label{tab: GWS_T_1314}
\begin{tabular}{lllllllrrrc}  
\headrow
\thead{\#} & \thead{GWS} & \thead{Height  (m a.s.l)} & \thead{} & \thead{} & \thead{ME} & \thead{}& \thead{} & \thead{} & \thead{RMSE} & \thead{}\\
\thead{T2} & \thead{} & \thead{} & \thead{YSU} & \thead{MYJ} & \thead{BouLac} & \thead{KEPS-TPE}& \thead{YSU} & \thead{MYJ} & \thead{BouLac} & \thead{KEPS-TPE}\\
\hiderowcolors
1 & BR & 226 & -2.4 & -3.8& \textbf{-1.1} & -3.0 & 2.6 & 3.8 & \textbf{1.3} & 3.2 \\
2 & BO & 254 & -2.7 & -4.0 & \textbf{-1.2} & -3.1 & 3.0 & 4.2 & \textbf{1.4} & 3.5\\
3 & GA & 290  & -1.9 & -2.6 & \textbf{-0.4} & -2.4 &  2.2 & 2.8 & \textbf{1.3} &2.7 \\
4 & ME & 330  & -1.5 & -2.7 & \textbf{-0.8} & -2.0 &  1.9 & 3.0 & \textbf{1.3} & 2.3\\
5 & BC & 490 & -1.0 & -3.0 & \textbf{-0.1} & -1.6 &  1.6 & 3.4 & \textbf{1.1} & 2.1\\
6 & CA & 495 & -1.5 & -3.1 & \textbf{-0.9} & -2.0 & 1.8 & 3.3 & \textbf{1.1} & 2.2 \\
7 & BS & 585 & 0.4   & -1.3 & 1.4 & \textbf{-0.3} &  \textbf{1.1}& 2.0 & 1.6 & 1.2 \\
8 & FS & 840 & -0.9 & -2.2 & \textbf{-0.2} & -1.4 & 1.1 & 2.3 & \textbf{0.6} & 1.5\\
9 & SA & 970 & -1.2& -2.2 & \textbf{-0.7} & -1.4 & 1.8 & 2.5 & \textbf{1.5} & 2.0 \\
10 & SG & 970 & -1.0 & -2.0 & \textbf{-0.6} & -1.2 &  1.4 & 2.2 & \textbf{1.2} & 1.6\\
11 & NP & 1470 & -0.5 & -2.0 & \textbf{0.3} & -0.8 & 1.7 & 2.3 & \textbf{1.6}& 2.0\\
& & \textbf{<ME> $\quad$ <RMSE>}& -1.3 & -2.6 & \textbf{-0.4} & -1.8 & 1.8 & 2.9 & \textbf{1.3} & 2.2 \\
\hline  
\end{tabular}
\end{table}

\begin{table}[bt]
\caption{ME and RMSE for the 10-m wind speed of the different simulations for Episode 2. ME and RMSE values are computed from hourly averages of GWS measurements and WRF outputs in the interval 13-02-2017 15:00 LST - 14-02-2017 18:00 LST. The symbol \textsuperscript{*} means $|\textrm{ME}|$ or RMSE less than 0.05. <ME> and <RMSE> refer to the average values of the metrics considering all the GWSs.}
\label{tab: GWS_WS_1314}
\begin{tabular}{lllllllrrrc}  
\headrow
\thead{\#} & \thead{GWS} & \thead{Height  (m a.s.l)} & \thead{} & \thead{} & \thead{ME} & \thead{}& \thead{} & \thead{} & \thead{RMSE} & \thead{}\\
\thead{WS10} & \thead{} & \thead{} & \thead{YSU} & \thead{MYJ} & \thead{BouLac} & \thead{KEPS-TPE}& \thead{YSU} & \thead{MYJ} & \thead{BouLac} & \thead{KEPS-TPE}\\
\hiderowcolors
1 & BR & 226 & -0.7& -0.7&\textbf{0.4}& -0.9& 1.5& 1.7& \textbf{1.0}&1.7\\
2 & BO & 254 & -1.0 & -1.6& \textbf{-0.8}& -1.0&2.1 & 2.3& 2.0&\textbf{1.8}\\
3 & GA & 290  & 0.4& 0.4 & 0.4& \textbf{0.3} &  0.7& 0.7& 0.9& 0.7\\
4 & ME & 330  & 0.5  & 0.4 & \textbf{0.3} & 0.4 &  1.3 & \textbf{0.7} &   1.2& 1.2\\
5 & BC & 490 & 1.0 & \textbf{0.2} & 1.2 &1.1&  1.9& \textbf{1.0}&2.3&1.6\\
6 & CA & 495 & -0.6& -0.6& \textbf{-0.1}& -0.4 &1.2&1.2& \textbf{1.0}&1.2\\
7 & BS & 585 & \textbf{-0.1}& -0.6& 0.6&-0.2 & 1.3& 1.2&1.3&\textbf{1.0}\\
8 & FS & 840 &  \textbf{0.0\textsuperscript{*}}& -0.2& 0.1 & 0.1&  0.6 & \textbf{0.5}& 0.8 &0.7\\
9 & SA & 970 & -0.1&-0.5& 0.1& -0.3 &  \textbf{0.9} & 1.0 &  1.1&1.1\\
10 & SG & 970 & 0.4& \textbf{0.1}& 0.5& 0.4&  0.7 & \textbf{0.4}&0.8&0.7\\
11 & NP & 1470 & 0.2& 0.1& 0.2& 0.1& 0.8 & 0.8& 1.0& 0.8\\
& & \textbf{<ME> $\quad$ <RMSE>}& 0.0* & -0.3 & 0.3 & -0.0* & 1.2 & \textbf{1.0} & 1.2 & 1.1 \\
\hline  
\end{tabular}
\end{table}
\subsubsection{Quantitative evaluation of model results}

As for Episode 1, we compare the ME and RMSE on an hourly basis, in order to quantitatively evaluate the model performance with the different PBL schemes. Results for the 11 GWSs are summarized in Table~\ref{tab: GWS_T_1314} and Table~\ref{tab: GWS_WS_1314} for 2-m temperature and 10-m wind speed, respectively. 

A general underestimation of temperature is observed ($\sim$ \SI{1}{\degreeCelsius} for all schemes, except for MYJ with <ME>=-\SI{2.6} {\degreeCelsius}), with BouLac having the lowest ME and RMSE across all GWSs except at Bressanone (BS). In particular, a strong underestimation is found at Bolzano (BO), which was not found in Episode 1. As mentioned in Sec. \ref{subsec: case studies}, Episode 2 was characterised by the presence of stratocumulus clouds over the Bolzano basin and the Adige Valley, which are likely underestimated by the model, leading to a stronger radiative cooling at the surface and a lower temperature at Bolzano. 
Also for this episode, the ME of wind speed is quite small (generally less than \SI{1}{\meter \per \second}), with little sensitivity to the PBL scheme. 
Considering the RMSE, MYJ shows the lowest value, although the variability between the different simulations is generally low. 

Figure~\ref{fig: GWS_1314_BBB} shows the hourly time series of 2-m temperature and 10-m wind speed at the GWSs selected for Episode 1. A systematic underestimation of temperature is observed for all PBL schemes, both during nighttime and daytime, with BouLac being the one closest to observations. 
In contrast to Episode 1 (cf. Fig.~\ref{fig: GWS_2829_BBB}), a constant wind speed of \SI{3}{\meter \per \second} is observed at Barbiano-Colma during the night. This feature is properly reproduced by the model, and the choice of the PBL scheme mostly influences the onset and ending time of this drainage flow. At the same time, an intense wind at ground level is observed at Bolzano, which was not present in Episode 1, when a ground-based temperature inversion was well-developed in the basin. This feature is captured by YSU and KEPS-TPE, and partly by BouLac, but with a shorter duration compared to observations. Therefore, the model with the BouLac scheme underestimates the penetration of the valley-exit wind at low levels in the Bolzano basin, possibly due to the overestimation of the temperature stratification (cf. Fig.~\ref{fig: Fig9_1314_profiler}d). On the other hand. MYJ is completely missing the nocturnal wind increase at Bolzano GWS, even if the wind speed and direction are generally well captured at the lidar location (Fig~\ref{fig: Fig10_1314_lidar}c and Fig~\ref{fig: Fig11_1314_lidar_dir}c).

At Bronzolo, the wind speed is very weak during nighttime, supporting that the wind observed at Bolzano is related to the drainage flow of the Isarco Valley. A peak in wind speed is observed during the afternoon, related to the strong up-valley flow of the Adige Valley, and correctly reproduced only by the BouLac scheme. This corroborates the hypothesis drawn from observations that a channeling mechanism was strengthening the up-valley flow in the Adige Valley. However, the comparison at Barbiano-Colma shows that all schemes, except MYJ, overestimate this effect in the Isarco Valley. This can explain the delay, compared to lidar observations, in the onset of the valley-exit wind, observed in all simulations (Fig.~\ref{fig: Fig10_1314_lidar}). 
\begin{figure}[h!]
\centering
\includegraphics[width=\linewidth]{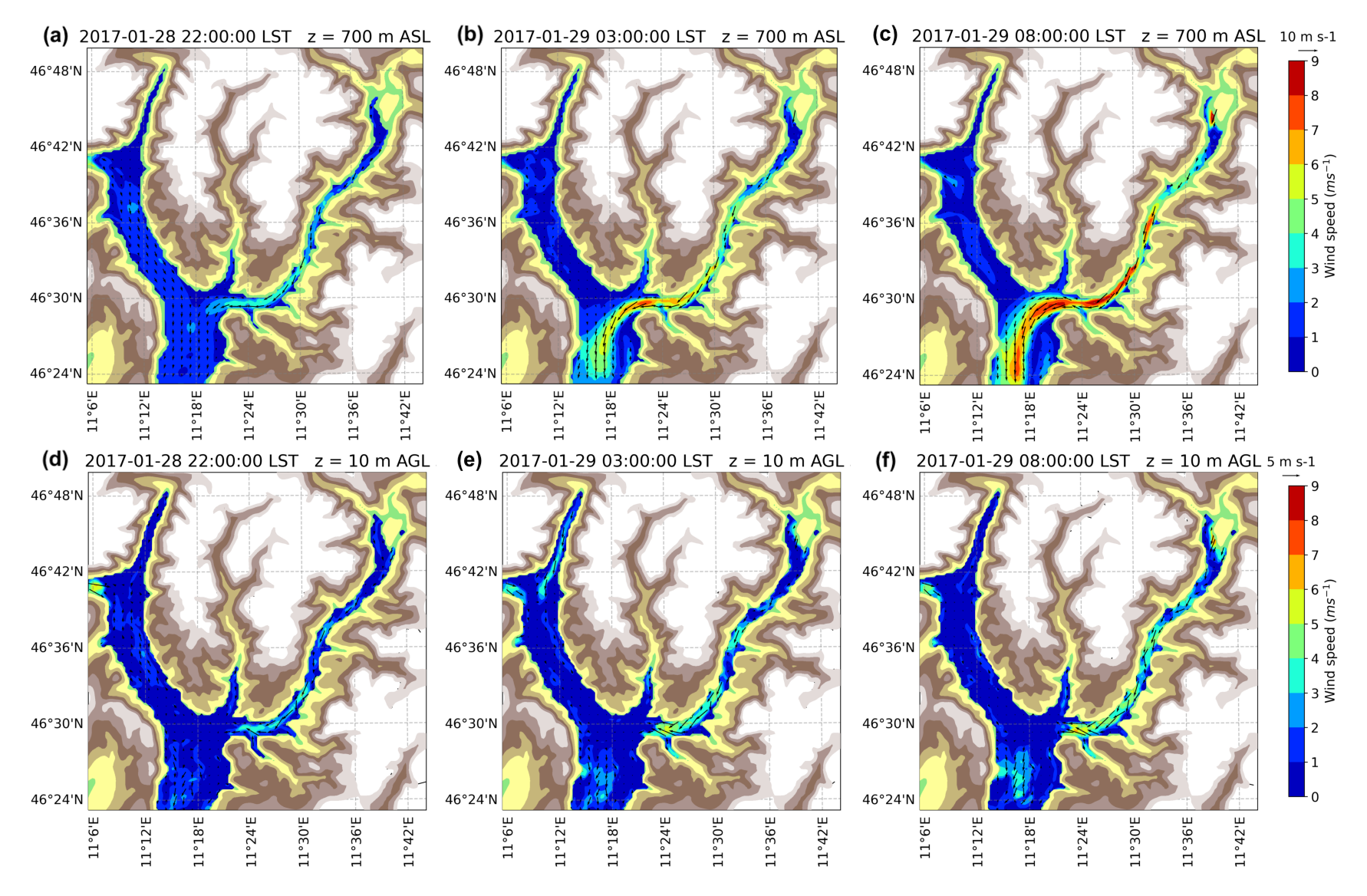}
\caption{Horizontal cross-sections of hourly wind speed at \SI{700}{\meter} a.s.l (a, b, c) and at \SI{10}{\meter} AGL (d, e, f) for Episode 1, as reproduced by the YSU simulation at (a, d) 22:00 LST 28 January 2017, (b, e) 03:00 LST 29 January 2017, (c, e) 08:00 LST 29 January 2017.}
\label{fig: 2829 Horizontal CS}
\end{figure}

\begin{figure}[h!]
\centering
  \includegraphics[width=0.4\linewidth]{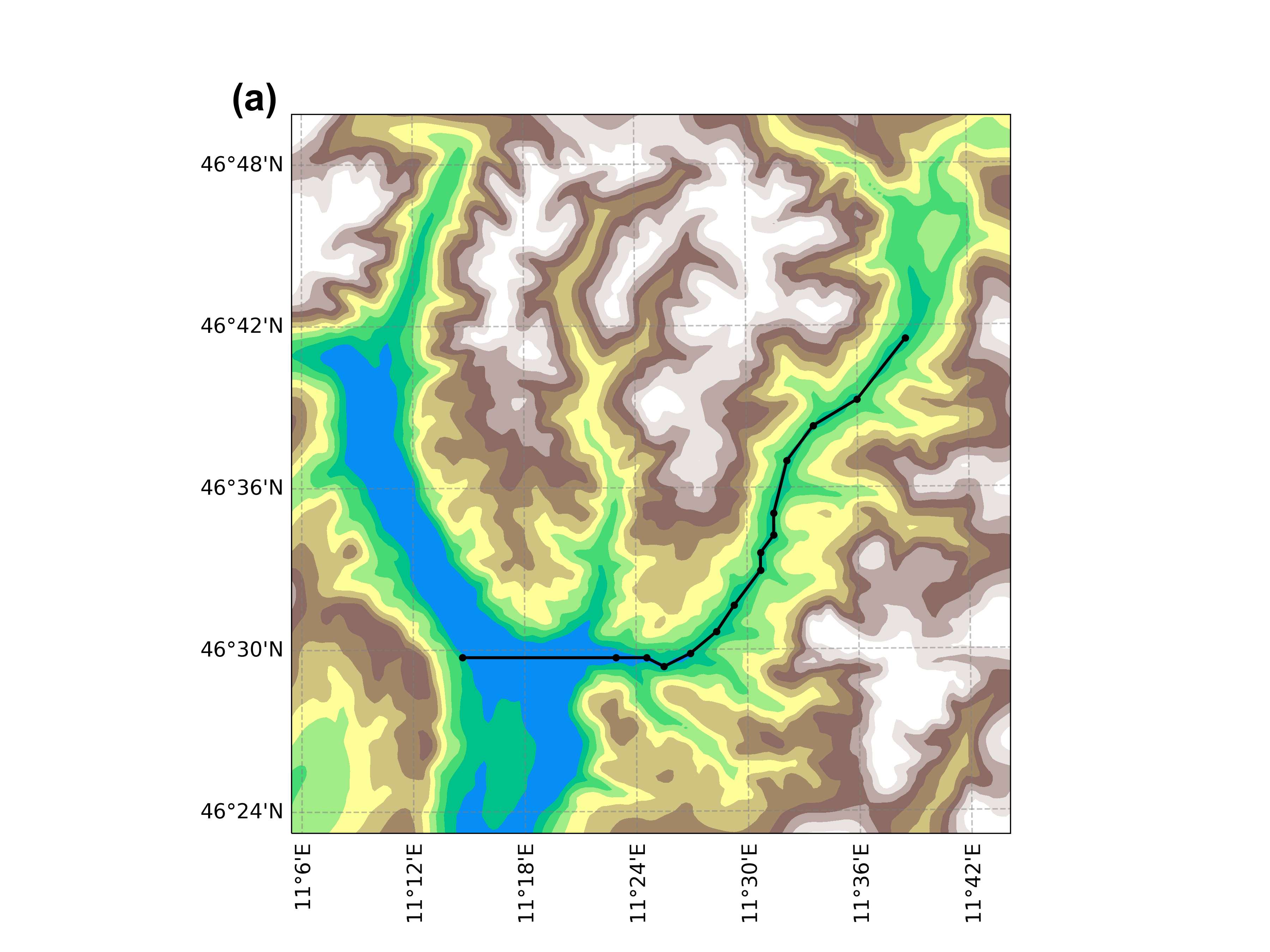}
  \par\smallskip
  \includegraphics[width=\linewidth]{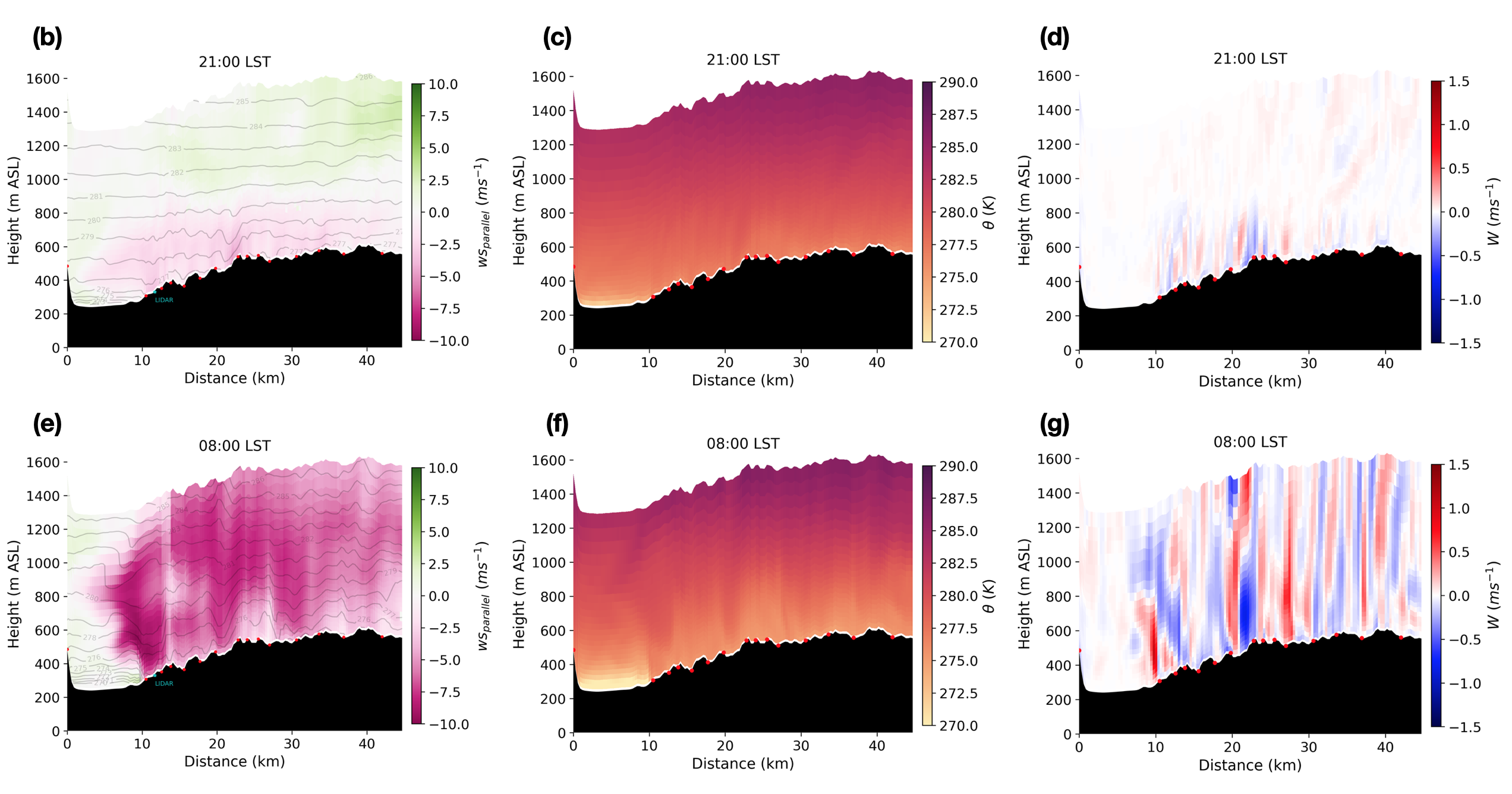}
\caption{Vertical cross-sections from the YSU simulation along the Isarco Valley, following the black line in panel (a), for Episode 1 at (a, b, c) 21:00 LST 28 January 2017, and (d, e, f) 08:00 LST 29 January 2017. Panels (b) and (e) show the wind speed parallel to the cross-section plane (color scale), with negative values for easterly (down-valley flow) and the potential temperature (gray lines), panels (c) and (f) show the potential temperature, panels (d) and (g) show the vertical wind speed. The terrain height is shown in black with the red points as delimiters between the different cross-sections (see panel (a)).}
\label{fig:2829VertCS}
\end{figure}

\section{Spatial characteristics of the valley-exit wind and interaction with the CAP}
\label{sec: interaction}
Numerical simulations provide a means to analyse the spatial characteristics of the valley-exit flow and the mechanisms governing its interaction with the Bolzano basin air, complementing the information from sparse point measurements. For this purpose, the YSU scheme was selected for Episode 1 and the KEPS-TPE scheme for Episode 2. These specific simulations were chosen due to their good agreement with the lidar observations regarding both horizontal (Figs.~\ref{fig: Fig5_lidar} and \ref{fig: Fig10_1314_lidar}) and vertical (Figs.~\ref{fig: Fig7_lidar_w} and \ref{fig: Fig12_1314_lidar_w}) wind velocity components. However, as the verification analysis revealed that no single scheme clearly outperforms the others, the findings presented in this section remain robust regardless of the specific simulation selected.

For Episode 1, Figure~\ref{fig: 2829 Horizontal CS} presents horizontal cross-sections of the wind field at different hours of the night at $\sim$\SI{700}{\meter} a.s.l. (top row) and \SI{10}{\meter} AGL (bottom row). At $\sim$700 m a.s.l. the down-valley flow is visible both in the Adige Valley and in the Isarco Valley (Fig.~\ref{fig: 2829 Horizontal CS}a). The modeled flow in the Isarco Valley shows that larger wind speeds are reached in the lower section of the valley, corroborating the hypothesis drawn from the observations that the nocturnal drainage current presents valley-exit jet characteristics. As night progresses, at 700 m a.s.l. the valley-exit flow intensifies and, once entering the basin, it veers southward before spreading horizontally in the Adige Valley (Figs.~\ref{fig: 2829 Horizontal CS}b,c). It veers southward before reaching the mountains in the westward direction, which may point to a peculiar effect of the valley geometry or to an interaction with the down-valley flow of the Adige Valley. 
Just beyond the valley exit, the 10-m wind speed abruptly decreases (Figs.~\ref{fig: 2829 Horizontal CS}e,f). This suggests that a well-developed CAP covers the basin at low levels and the drainage flow is not able to split, e.g., flowing partly above and partly below the CAP, as observed in \citet{Cuxart2007}, or flowing around it. The CAP blocks the valley-exit flow from penetrating at low levels and the entire drainage flow is forced to lift over the colder, denser air within the basin. 

Figure ~\ref{fig:2829VertCS} shows vertical cross sections along the Isarco Valley displaying the horizontal wind speed parallel to the section (Figs.~\ref{fig:2829VertCS}b,e), potential temperature (Figs.~\ref{fig:2829VertCS}c,f) and vertical wind speed (Figs.~\ref{fig:2829VertCS}d,g), along the black line shown in Fig.~\ref{fig:2829VertCS}a. This transect was chosen to follow the topography along the main axis of the Isarco Valley and to capture the transition to the Bolzano basin. In the first hours of the night, the down-valley flow (corresponding to negative values in Fig.~\ref{fig:2829VertCS}b) is weak, shallow, and uniformly developed. Throughout the night, its depth increases across the whole Isarco Valley with stronger speeds at the exit region (Fig.~\ref{fig:2829VertCS}e). Here, the flow sinks and compresses, confirming that its intensification is due to mechanisms similar to those leading to the formation of valley-exit winds. The sharp flow deceleration at the exit region extends across multiple vertical layers and the isentropic lines show that it is connected to the depth of the CAP in the basin.
The cross-sections of potential temperature (Figs.~\ref{fig:2829VertCS}c,f) show that a CAP is already present at the beginning of the night, covering the whole basin. Conversely, the Isarco Valley cools uniformly but maintains a higher potential temperature than the basin. In the morning, a strong CAP still persists, with a sharp boundary at the valley exit. Consistent with lidar observations (Fig.~\ref{fig: Fig12_1314_lidar_w}a), the model reproduces upward velocities from the surface up to $\sim$800 m a.s.l. at the valley exit (Fig.~\ref{fig:2829VertCS}g). This suggests that the CAP is acting as a downstream block to the flow \citep{Reeves2006}. 

Downward velocities just above the rising layer are simulated, suggesting that the valley-exit wind rises above the CAP while it is still sinking and compressing, which may contribute to further intensifying the flow. 

\begin{figure}[h!]
\centering
\includegraphics[width=\linewidth]{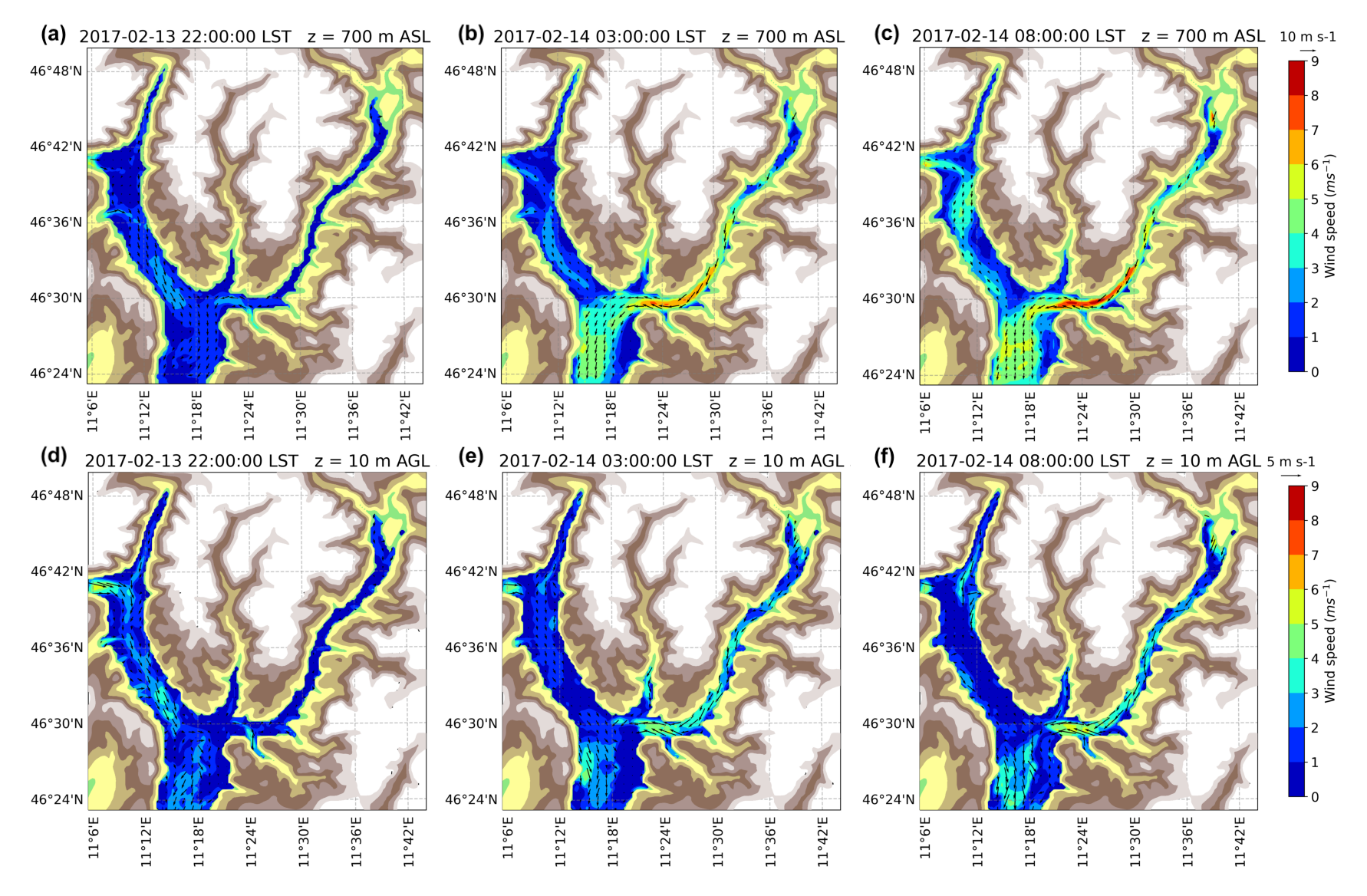}
\caption{Horizontal cross-sections of hourly wind speed at \SI{700}{\meter} a.s.l. (a, b, c) and at \SI{10}{\meter} AGL (d, e, f) for Episode 2, as reproduced by the KEPS-TPE simulation at (a, d) 22:00 LST 13 February 2017, (b, e) 03:00 LST 14 February 2017, (c, e) 08:00 LST 14 February 2017.}
\label{fig:1314VertCS Horizontal CS}
\end{figure}

\begin{figure}[t]
\centering
\includegraphics[width=\linewidth]{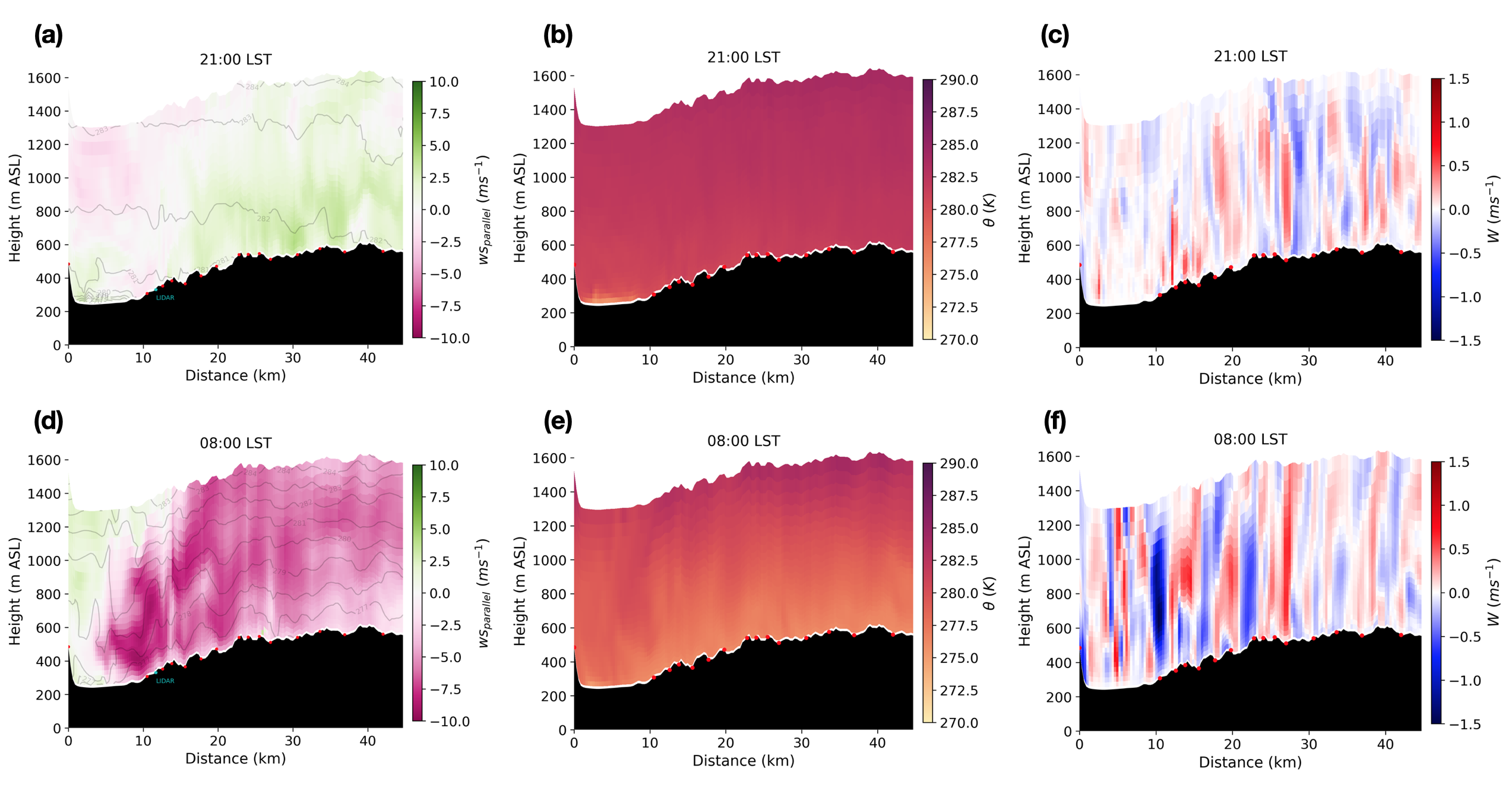}
\caption{Vertical cross-sections from the KEPS-TPE simulation along the Isarco Valley, following the black line in Fig.~\ref{fig:2829VertCS}a, for Episode~2 at (a--c) 21:00~LST 13 February 2017 and (d--f) 08:00~LST 14 February 2017. Panels (a) and (d) show the wind speed parallel to the cross-section plane (negative values indicate down-valley flow), together with potential temperature contours (gray). Panels (b) and (e) show the wind speed parallel to the cross-section plane (color scale) with potential temperature (gray contours). Panels (c) and (f) show potential temperature. The terrain height is shown in black; red points delimit the different cross-sections (see Fig.~\ref{fig:2829VertCS}a).}
\label{fig:1314VertCS}
\end{figure}

For Episode 2, the horizontal wind field at $\sim$\SI{700}{\meter} a.s.l., shown in Figs.~\ref{fig:1314VertCS Horizontal CS}a,b,c, exhibits a pattern similar to Episode 1 (cf. Figs.~\ref{fig: 2829 Horizontal CS}a, b, c). However, the flow tends to be weaker, and spreads more broadly across the Adige Valley, which is likely a consequence of its weaker intensity. Near the surface, wind speeds similar to Episode 1 are reproduced by the model at the valley exit, but, in this case, the down-valley flow propagates further into the basin. This spatial structure of the flow is consistent with the observations from the Bolzano GWS, which measured near-surface winds of about \SI{4}{\meter \per \second} for Episode 2 (Fig.~\ref{fig: GWS_1314_BBB}), in contrast with the calm conditions for Episode 1 (Fig.~\ref{fig: GWS_2829_BBB}). Vertical-cross sections analogous to Fig.~\ref{fig:2829VertCS} are shown in Fig.~\ref{fig:1314VertCS}. The onset of the down-valley flow in the Isarco Valley occurs later than in Episode 1, a delay attributed to the southerly synoptic wind channeling mentioned in Sec.~\ref{sec: Results_ep2}. The atmosphere is less stratified, as evidenced by the wider spacing of the isotherms in Figs.~\ref{fig:1314VertCS}a,d compared to Episode 1 (Figs.~\ref{fig:2829VertCS}a,d). By early morning, the down-valley flow has reached depths similar to Episode 1 and, again, it sinks and compresses at the exit, confirming the hydraulic flow transition as the main dynamical mechanism for the flow intensification. However, in this case the flow is able to penetrate into the Bolzano basin down to the lowest layers because the valley-exit wind has a lower potential temperature than the air in the basin (Fig.~\ref{fig:1314VertCS}e). Accordingly, the exit region is characterised by a downward motion even at the ground (Fig.~\ref{fig:1314VertCS}f) with the flow following the topography toward the basin.

\section{Conclusions}
\label{sec: conclusions}
An intense field campaign took place in a mountain basin in the Adige Valley (Italian Alps) during winter 2017, including surface measurements and vertical profiles of temperature and wind speed and direction. Wind profile measurements were performed near the exit of a tributary valley into the basin, and strong wind speeds around \SI{10}{\meter \per \second} were observed at night, with maximum intensity close to sunrise, pointing to a strong drainage flow from the tributary valley. Similar speeds in mountainous environments have been observed at regions where topography abruptly changes (e.g., canyon and valley exits) and attributed to the sinking and compression of a cold-air flow at those regions, with a consequent acceleration of the flow. These flows are commonly named valley-exit jet/flows and the present study aimed to investigate the characteristics of the drainage current into the basin, and understand whether the flow behaves as a valley-exit flow. Combining the information from wind and temperature measurements, observations also showed that the thermal stratification in the basin influences the flow once it reaches the basin. This feature was investigated in the present study by comparing the nocturnal circulation in two different wintertime cases: one with a strong ground-based temperature inversion, which led to the development of a CAP during nighttime, and one without.

The sparse information from point measurements was complemented with high-resolution ($\Delta \sim$ \SI{300}{\meter}) numerical simulations, performed with the WRF model. Model results are used to understand the main mechanism for the flow intensification at the valley exit, as well as the interaction of the flow with the CAP. Due to the challenges of numerical models to represent the very stable boundary layer, the choice of the PBL scheme may play a significant role in the correct representation of this nocturnal circulation. Thus, a comparison of simulations with different PBL schemes was performed to assess their possible impact on the model results and ensure robustness of the conclusions. Four PBL schemes (YSU, MYJ, BouLac, KEPS-TPE) have been tested and compared to measurements from a temperature profiler, a Doppler wind lidar, and 2-m temperature and 10-m wind speed from several ground weather stations on the valley floor and the sidewalls.

Simulation results indicate that the model is able to capture the main structure of the wind flow at the lidar location, quite independently of the PBL scheme, whereas the thermal structure at the thermal profiler location is more sensitive to the specific scheme. When a strong CAP develops during nighttime, the thermal stratification is best captured by a PBL parameterisation including, among others, a prognostic equation for temperature variance and a counter-gradient term (KEPS-TPE, \citealp{Zonato2022}). 

Cross sections of the wind field from the simulations confirm that the drainage flow has valley-exit wind characteristics: it sinks and compresses when approaching the exit of the valley with an increase in the wind speed. Cross sections of potential temperature also suggest that the CAP covers the entire basin such that the valley-exit wind rises as it encounters the cold air inside the basin, instead of splitting above and below the CAP, or flowing around it. 

When the thermal stratification in the basin is weaker, all PBL schemes reproduce quite well the vertical structure of the lower atmospheric layers, with an overall underestimation of temperature by 2-3 \textdegree C. This was likely caused by an inadequate representation of the stratocumulus clouds over the Adige Valley that were present in this case. All simulations produce a delay in the onset of the valley-exit wind, which was attributed to a strong up-valley wind in the afternoon lasting until the evening. Observations suggest a channeling mechanism, induced by the synoptic flow, in the Adige Valley, but the model tends to reproduce and strengthen this effect also in the Isarco Valley. 

During the night, a downward motion at the lidar location and strong 10-m wind speed at the GWSs along the path of the valley-exit wind are observed. This suggests that, with a weaker thermal stratification, the flow is able to reach the low layers in the basin, as also supported by cross sections of wind speed from the numerical simulations. 

This work provides a broad picture of a peculiar phenomenon of the local circulation in the Adige Valley. However, analogous situations may be found in other regions. Similar flows are reported at the exit of narrow valleys onto wider adjacent areas \citep{Zangl2004,Jimenez2019,Pfister2024} and drainage flow interactions with cold air inside a main valley \citep{Neff1987,Neff1989,Pinto2006,Munoz2020} are observed in different locations worldwide. Thus, results from this work may be useful to interpret the mechanisms and factors controlling similar circulations at other locations. 

Future investigations aim at better understanding the impact of the valley-exit flow on turbulent vertical mixing over the area. \cite{Lampert2016} suggests that the occurrence of a nocturnal low-level jet influences the anisotropic characteristics of turbulence in the stable boundary layer, by enhancing the damping of the vertical wind variance with respect to horizontal wind variances (on scales of a few hundred metres) 
and it would be interesting to investigate if these flows can drive turbulence towards a particular state of anisotropy of the Reynolds stress tensor, as observed for waves in stable stratified atmosphere \citep{Gucci2022}. Comparing cases with and without the arrival of the valley-exit flow in the basin could be used for this purpose. However, a new ad-hoc field campaign including turbulence measurements would be required.

\section*{ACKNOWLEDGEMENTS}
The authors acknowledge Eco–Center s.p.a. for the financial support to the project, in particular for renting the Doppler wind lidar. The Weather Service of the Autonomous Province of Bolzano provided data from its weather stations. The Environmental Agency of the Autonomous Province of Bolzano provided data from the microwave temperature profiler.

\section*{CONFLICT OF INTEREST}
The authors declare no conflict of interest.
\printendnotes

\section*{DATA AVAILABILITY STATEMENT}
The data that support the findings of this study are available from the corresponding authors upon request.
\bibliography{library2}

\end{document}